\documentclass[twocolumn,secnumarabic,amssymb,amsmath,nofootinbib,tightenlines,nobibnotes,aps,prb]{revtex4}
\usepackage{pict2e}
\usepackage{amsfonts}
\usepackage{mathtools}
\usepackage[usenames,dvipsnames]{color}
\usepackage{hyperref}
\usepackage{graphicx}
\usepackage{subfigure}

\newcommand{\qe}{\tilde\varepsilon} 
\newcommand{\nl}{\smallskip\newline\noindent}
\newcommand{\fvec}{\mathbf{f}}
\newcommand{\uu}{\hat{\mathcal{U}}}
\newcommand{\ww}{\hat{\mathcal{W}}}
\newcommand{\wwb}{\hat{\mathbb{W}}}
\newcommand{\g}{\mathcal{G}}
\newcommand{\istar}{\mathcal{I^{*}}}
\newcommand{\tp}{t_{\perp}}

\def\frac#1#2{{\textstyle{#1 \over #2}}}
\newcommand{\yd}{^\dagger}
\newcommand{\nd}{^{\vphantom{\dagger}}}
\newcommand{\va}{\vec a}
\newcommand{\half}{\frac{1}{2}}
\newcommand{\kperp}{\vec k\nd_{\perp}} \newcommand{\kpara}{k\nd_{\parallel}}

\def\EF{E_\textsf{F}}
\def\wwp{{\raise2.5pt\hbox{$\wp$}}}

\begin{document}

\title{Edge States, Entanglement Spectra, and Wannier Functions in Haldane's
Honeycomb Lattice Model and its Bilayer Generalization}
\author{Zhoushen Huang}
\author{Daniel P. Arovas}
\affiliation{Department of Physics, University of California, San Diego}
\date{\today}
\begin{abstract}
  We study Haldane's honeycomb lattice model and a bilayer generalization thereof
  from the perspective of edge states, entanglement spectra, and Wannier function behavior.
  For the monolayer model, we obtain the zigzag edge states analytically, and identify the edge state
  crossing point $k_{\rm c}$ with where the $f=\half$ entanglement occupancy
  and the half-odd-integer Wannier centers occur.  A continuous interpolation between the
  entanglement states and the Wannier states is introduced. We then construct a bilayer model by
  Bernal stacking two monolayers coupled by interlayer hopping.  We analyze a particular limit
  of this model where an extended symmetry, related to inversion, is present.  The band topology
  then reveals a break-down of the correspondence between edge and entanglement spectrum, which
  is analyzed in detail, and compared with the inversion-symmetric $\mathbb{Z}_2$
  topological insulators which show a similar phenomenon.
\end{abstract}

\maketitle

\section{Introduction}
Haldane's honeycomb lattice model\cite{haldane88} has provided a fertile
paradigm for topological band structures in the absence of net magnetic flux.
Prior to Haldane's work, Thouless {\it et al.\/}\cite{tknn82} (TKNN) showed how a tight binding
model with uniform rational flux per plaquette, {\it i.e.\/} the Hofstadter model,
yields a topological band structure in which each energy band $n$ is classified
by an integer topological invariant $C_n$, known in mathematical parlance as a Chern number.
(In the continuum limit, where the flux per plaquette is infinitesimal, the TKNN bands
become dispersionless Landau levels.)  The essence of the Haldane model lies in its
inclusion of complex second-neighbor hopping amplitudes, so that the model breaks
time-reversal ($\mathcal{T}$) symmetry even though the net magnetic flux per plaquette
vanishes, which allows for the existence of topological phase with band Chern indices $\pm 1$.

One of their hallmarks of topological phases is the
existence of gapless edge modes interpolating the bulk gap in the
presence of an open boundary. The number of such edge spectral flows
as functions of the momentum parallel to the boundary is the same as
the total Chern index of the bands below the gap, as first elucidated
by Hatsugai\cite{hatsugai93}. Kane and Mele \cite{kane-mele05-qsh}
later generalized Haldane's model by introducing spin and treating the
(now purely imaginary) second neighbor hopping as a spin-orbit
coupling. $\mathcal{T}$-preserving perturbations are also allowed. The
Kane-Mele model is $\mathcal{T}$-invariant and cannot have a quantum Hall
effect. The bulk topological property is instead described by a
$\mathbb{Z}_2$ topological index \cite{kane-mele05-z2,fu06-trz2}.
Remarkably, at half filling, while the total Chern index is zero, the
gapless edge spectrum persists due to time-reversal symmetry. A
topologically trivial band insulator, by contrast, would have no
edge spectral flow at all.

The gapless edge spectral flow of topological insulators is one of the
real space manifestations of their bulk topology. A similar spectral
flow can be observed in the quantum entanglement spectrum of the
many-body reduced density matrix obtained by partitioning the system
along a translationally-invariant boundary
\cite{Li-Haldane08,Haldane09}. For noninteracting fermions, the
spectrum of the reduced density matrix itself corresponds to that of a
noninteracting `entanglement Hamiltonian' determined by the one-body
correlation matrix of the original system \cite{cheong04,peschel03}.
There are however exceptions to the entanglement and edge spectra
correspondence. For example, the entanglement spectrum has protected
midgap modes for systems with inversion ($\mathcal{I}$) symmetry even
if the edge spectrum is gapped \cite{turner10-inversion,hughes-prodan-bernevig11-inversion}
by \emph{e.g.} breaking the $\mathcal{T}$ symmetry in quantum spin Hall effect (QSH) systems.
In certain cases, one also has to tune the boundary conditions for a system with nontrivial topology
in order for its energy edge modes to be gapless \cite{qi-wu-zhang06-tbc},
while such tuning is not required to observe the entanglement spectral flow.
We shall see similar differences in our study of the Haldane models. Thus in certain sense, the
entanglement spectrum reveals the bulk topology better than the Hamiltonian's edge spectrum.

The band topology can also be considered from a Wannier function \cite{kohn59-wannier} point of view.
While in higher dimensions, the construction of exponentially
localized Wannier functions is precluded for band insulators with
nonzero Chern numbers \cite{brouder07-wannier-obstruction}, the system
is effectively one-dimensional if one specifies the momentum $\kperp$
along the edge/surface, for which the Wannier states are well defined.
The Wannier functions are then localized along strips or planes parallel
to the edge.  Several recent studies of topological insulators
\cite{Yu11-z2-wannier, soluyanov-vanderbilt11-z2-wannier} have invoked
the Wannier states in their analysis. The Wannier centers are shown to
exhibit a spectral flow similar to that of the entanglement spectrum,
and the topological information can be visually extracted from their
flow pattern. Mathematically, the deviation of Wannier centers from
the corresponding unit cells are eigenvalues of the Wilson loop
operator $\ww$ \cite{Qi11-wannier}. It is interesting that $\ww$ is an
object derived purely from the bulk (for translationally invariant
systems) and is hence faster to compute, yet its eigenvalues have a
real space interpretation similar to the entanglement spectrum: when
the Wannier centers migrate from one unit cell to its neighbor, there
is a corresponding flow in the entanglement spectrum if the particular
unit cell boundary is used as the entanglement cut. The entanglement
spectrum is thus a coarse graining of the Wannier centers with an
emphasis on the real space behavior near the entanglement cut
\cite{ha11-hofstadter}.

In this paper, we study the Haldane honeycomb lattice model and a
bilayer generalization thereof, both from a real space perspective,
combining the analysis of Hamiltonian edge states,
entanglement spectra and Wannier center flow. We first present an
analytical solution of the monolayer zigzag edge modes, identifying
the $k$ point where the the two edge modes cross. This plays the
role of one of the $\mathcal{T}$-invariant $k$ points in the QSH
models. We find that at the same $k$ point, there is an entanglement
occupancy mode fixed at $f=\frac{1}{2}$, and the corresponding Wannier centers
reside exactly in the middle of two neighboring unit cells. We show
that a common origin underlies this coincidence. We then extend to a
bilayer model by Bernal-stacking two monolayers with vertical
interlayer hopping, as in bilayer graphene. With a particular parameter choice,
the bilayer model exhibits something similar to the $\mathbb{Z}_2$
inversion-symmetric topological insulators (ITI) in that the edge
spectrum is gapped yet the entanglement spectrum has protected $f =
\frac{1}{2}$ modes. However, it cannot be explained in the ITI
framework due to the lack of $\mathcal{I}$ symmetry (the protected
modes do not occur at the inversion-symmetric $k$ points). We show
that the nontrivial topology is a consequence of a related $\istar$ symmetry
to be detailed in the text, and derive an expression to compute the
topological index, which is the winding number of one branch of the
Wannier centers. We further confirm this numerically by adding in
$\istar$-preserving perturbations to both the bilayer model and an ITI
model studied in Ref.~\onlinecite{hughes-prodan-bernevig11-inversion}.

\section{Monolayer Haldane Model}
\begin{figure}[htb]
  \centering
  \includegraphics[width=0.4\textwidth]{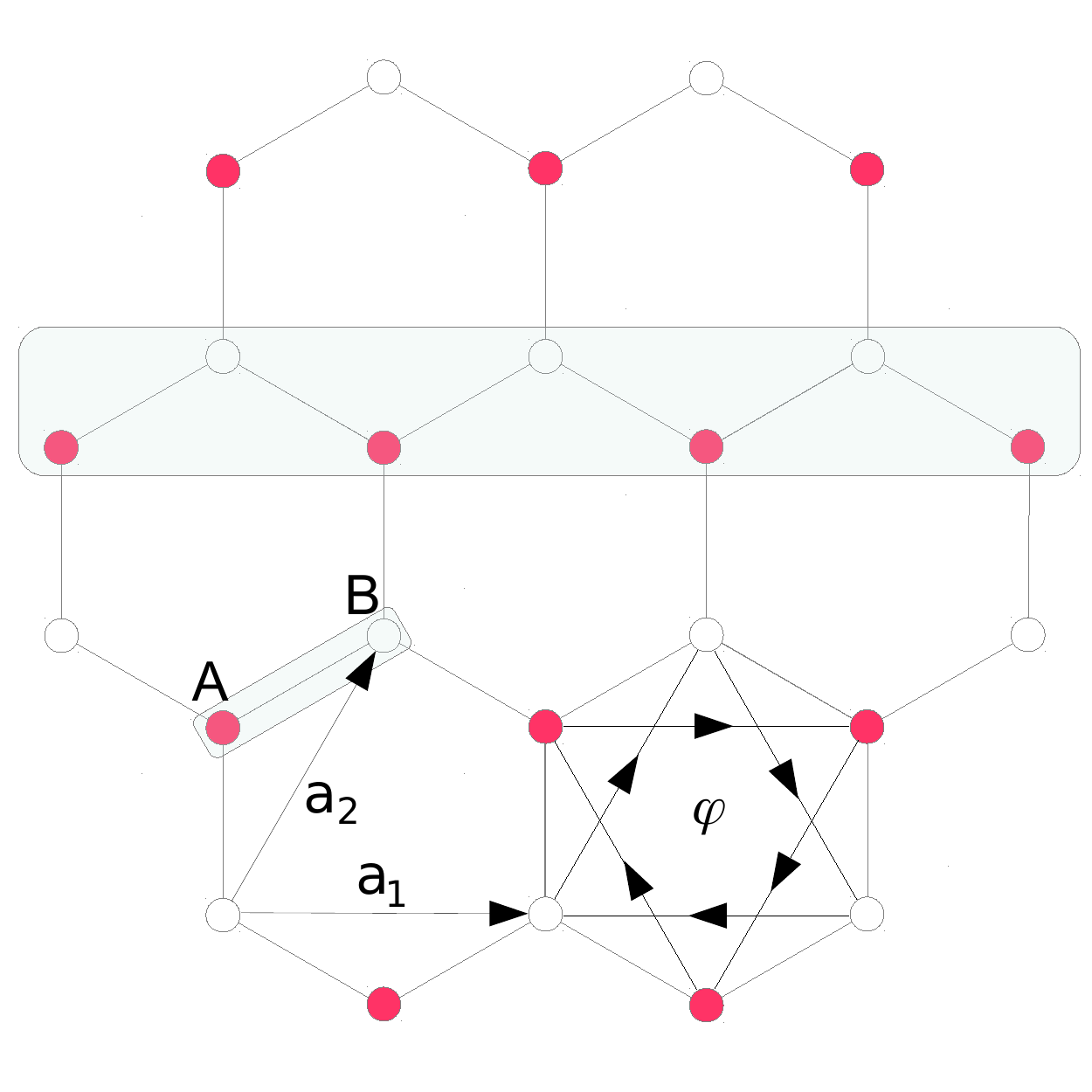}
  \caption{(Color online) Haldane's model. $A$ (red) and $B$ (white) label
    sublattice sites, and the boxed pair represents a unit cell.
    Primitive lattice vectors are chosen to be $\vec a_1$ and $\vec
    a_2$ as shown. Second neighbor hopping between same sublattice
    sites picks up a phase of $+\phi$ along the arrowed directions and
    $-\phi$ in opposite directions. Horizontal box encloses a zigzag
    edge.}
  \label{mlh-zz}
\end{figure}

We first briefly review the Haldane model, which is a tight binding model
on the honeycomb lattice, described by the Hamiltonian
\begin{gather}
  \mathcal{H} = -\sum_{i,j} t\nd_{ij} \, c\yd_i \, c\nd_j
 + \sum_i m\nd_i \, c\yd_i c\nd_i.
\end{gather}
The hopping amplitudes $t\nd_{ij}$ are nonzero only for nearest neighbor (NN)
and next-nearest neighbor (NNN) hopping.  For NN hops, $t_{ij}\equiv t$ is real.
For NNN hops, $t_{ij}=t_s\,e^{\pm i\phi}$, where $s=1$ if the hopping is parallel
to $\va_1$, $s=2$ if parallel to $\va_2$, and $s=3$ if parallel to
$\va_3\equiv \va_2-\va_1$. The sign of the phase is according to the arrows in
Fig. \ref{mlh-zz}, and is taken to be positive for clockwise hops within each hexagonal
unit cell.  In Haldane's original model, $t_1 = t_2 = t_3$.
Setting these amplitudes to be different breaks the three-fold rotational symmetry. The
Semenoff mass $m_i$ is $m$ and $-m$ for $A$ and $B$ sublattices
respectively, which breaks inversion.

In the bulk, the Fourier transformed Hamiltonian is
\begin{gather}
  H(\vec k) = \omega(\vec k) + \vec B(\vec k) \cdot \vec \sigma,
  \label{hbulk}
\end{gather}
$\vec \sigma = (\sigma_x, \sigma_y, \sigma_z)$ are the Pauli matrices
in the ``isospin'' degree of freedom, where $A$ and $B$ are isospin up
and down, respectively, and
\begin{align}
B_x &= -1 - \cos k_1 - \cos k_2 \label{omega-b} \\
B_y &= -\sin k_1 - \sin k_2 \notag \\
B_z &= m + 2\sin\phi \bigl[ t_1 \sin k_1 - t_2 \sin k_2 + t_3 \sin(k_2 - k_1) \bigr]  \notag\\
\omega &= -2 \cos\phi \bigl[ t_1 \cos k_1 + t_2 \cos k_2 + t_3\cos(k_2 - k_1) \bigr]  \notag 
\end{align}
Here, $k_i = \vec k \cdot \vec a_i \in [0, 2\pi]$ are the Bloch phases along
the two primitive vectors. The bulk topology is characterized by the
Chern number $C$ of the upper band, which is the winding number of the unit
vector ${\hat B}({\hat k})$ over the Brillouin zone.  That is to say, if by varying $(k_1, k_2)$
over the first Brillouin zone, $\hat B = \vec B/|B|$ covers the unit sphere once ($|C|=1$),
then the system is in its topological phase, otherwise it is in its trivial phase. 
Equivalently, in the topological phase the origin is inside the surface swept out by $\vec B$,
while in the nontopological phase it lies outside.  The topological phase
transition thus takes place when the origin is crossed by $\vec B$ at some
$\vec k$ points, where the gap $\Delta = 2|B|$ will vanish. Following
eqn.~\ref{omega-b}, this can only happen at the graphene Dirac points,
$(k_1, k_2) = \pm (2\pi/3, -2\pi/3) \equiv \vec K_{\pm}$, where $B_x =
B_y = 0$. The corresponding $B_z$ values necessarily have opposite
signs in the topological phase (so that the origin is enclosed),
\emph{i.e.}\cite{haldane88},
\begin{gather}
  \label{phase}
  |m| - \sqrt{3} \left|(t_1 + t_2 + t_3) \sin\phi\right|
  \begin{cases}
    < 0 & \text{topological}\\
    > 0 & \text{trivial}
  \end{cases}.
\end{gather}
In the trivial phase, the eigenstates just below the gap at both Dirac points are entirely concentrated on
the same sublattice.  In the topological phase, however, they are concentrated on opposite sublattices.

\subsection{Zigzag edge states}
\label{mlh-zzedge}

One way of solving the edge spectrum of tight binding models is to use
the transfer matrix formalism, following Hatsugai's investigation
\cite{hatsugai93} of the Hofstadter problem \cite{hof76}.  It is
worthwhile to first think about its application in the Haldane model.  In the
Hofstadter problem, the system is immersed in a uniform magnetic field
with rational flux $2\pi p/q$ per (square, say) plaquette, and the lattice vector
potential is periodic on the scale of the magnetic unit cell, which consists of
$q$ structural cells.  As a result, a $qN$-step transfer matrix
$\mathcal{M}_{qN}$ can be broken up into a product of $N$ identical
matrices each equal to a $q$-step transfer matrix, $\mathcal{M}_{qN} =
(\mathcal{M}_q)^N$, thus the solutions of $\mathcal{M}_q$ comprise a
special set of the solutions of $\mathcal{M}_{qN}$.  Physically this
means the \emph{edge} spectrum of a system with $qN-1$ structural cells in the
open direction is identical to the \emph{full} spectrum of a system
with $q-1$ structural cells.  Thus numerically, one only needs to diagonalize a
$(q-1)\times (q-1)$ Hamiltonian to find the edge spectrum of a $(qN-1)
\times (qN-1)$ Hamiltanion.  Clearly the method is most efficient in a
situation where the transfer matrix has such a decomposition,
\emph{e.g.}, a graphene sheet in magnetic field
\cite{hatsugai06-graphene-edge}. The Haldane model has no macroscopic
magnetic field (an essential virtue of the model), thus the transfer
matrix formalism yields little advantage over directly diagonalizing
the full Hamiltonian.  Still, one may employ it to analyze the Riemann
sheet structure of the complex energy, which is studied in
Ref.~\onlinecite{hao08}, but that is not our focus here.

A useful feature of Hatsugai's solution is that it
can be written as a direct product of an $N$-component real space part,
corresponding to the magnetic unit cell coordinate, and a $q$-component
internal space part, corresponding to the lattice points within each magnetic unit cell.
(See, {\it e.g.\/}, the appendix of Ref.~\onlinecite{ha11-hofstadter}.)  This is by no means
a general form for edge states.  All Bloch states on the other hand have
such a decomposition.  What we found for the Haldane model is that
the edge states in the case of a zigzag edge can also be direct-product-decomposed,
with the following caveats.  First, while the real space part in the Hofstadter model
has exact exponential dependence on the coordinate (equivalent to an
imaginary Bloch wavevector), which is due to decomposition of the
transfer matrix, this is not so in the Haldane model (nor is this surprising
since the macroscopic magnetic field is zero).  Second, the boundary
condition used in the Hofstadter model corresponds to an open edge. In
the Haldane model, the boundary condition must be \emph{tuned}
self-consistently to conform with the direct product {\it Ansatz\/}. Only
those boundary conditions with vanishing magnitude will correspond to
an edge state at an \emph{open} boundary, as opposed to, say, an
enhanced-tunneling boundary.

We now proceed to solve for the zigzag edge states.

\subsubsection{Twisted-boundary Hamiltonian and gauge transformation}

The zigzag edge is parallel to $\vec a_1$ (see the horizontal box in
Fig.~\ref{mlh-zz}), hence $k_1 = \vec k \cdot \vec a_1$ is a good
quantum number. Assume there are $N$ unit cells in the $\vec a_2$
direction. At each $k_1$, the effective $1$-D system is described by
the following Hamiltonian,
\begin{multline}
\label{h-tbc}
\mathcal{H}(k_1) = \sum_{n,n' = 1}^N\Bigl[
a_n\yd K^{(1)}_{n,n'} a\nd_{n'} + b_n\yd K^{(2)}_{n,n'} b\nd_{n'} \\
+ a_n\yd R\nd_{n,n'} b\nd_{n'} + b_n\yd R\yd_{n,n'} a\nd_{n'} \Bigr]
\end{multline}
Here $a\yd_n$ and $b\yd_n$ are creation operators on the A and B site of the
$n^{\rm th}$ unit cell, respectively.  The coefficient matrices $K^{(i)}$
connect sites on the same sublattice, and $R$ connects different
sublattices.  Their nonzero matrix elements are given by
\begin{align}
  \label{kimat}
  K^{(i)} &=
  \begin{pmatrix}
    h_i & v_i & & & \tilde v_{i}^{*}\\
    v_i^{*} & h_i & v_i \\
    & \ddots & \ddots & \ddots\\
    & & v_i^{*} & h_i & v_i\\
    \tilde v_{i}& & & v_i^{*} & h_i
  \end{pmatrix}\ ,
\end{align}
\begin{align}
  \label{rmat}
  R &= \begin{pmatrix}
    p_1 &  & & & \tilde v_{21}^{*}\\
    p_2 & p_1\\
    & \ddots & \ddots\\
    & & p_2 & p_1\\
    \tilde v_{12}& & & p_2 & p_1
  \end{pmatrix}\ .
\end{align}
The individual matrix elements can be obtained from Fourier
transforming the bulk Hamiltonian eqn.~\ref{hbulk},
\begin{gather}
  h \equiv \sum_{k_2} H(k_1, k_2) =
  \begin{pmatrix}
    h_1 & p_1\\ p_1^{*} & h_2
  \end{pmatrix}, \\
  \label{vmat}
  v \equiv \sum_{k_2} H(k_1, k_2) e^{-ik_2} =
  \begin{pmatrix}
    v_1 & 0 \\ p_2 & v_2
  \end{pmatrix}
\end{gather}
with 
\begin{align}
  h_1 &= m - 2t_1 \cos(\phi + k_1)\ ,\\
  h_2 &= -m - 2t_1\cos(\phi - k_1)\ ,\\
  v_1 &= -t_2 e^{-i\phi} - t_3e^{i\phi}e^{-ik_1}\ ,\\
  v_2 &= -t_2 e^{i\phi} - t_3e^{-i\phi}e^{-ik_1}\ ,\\
  p_1 &= -1 - e^{-ik_1} \quad, \quad p_2 = -1\ .
\end{align}
We note that swapping subscripts $1$ and $2$ on the left hand sides
yields a Hamiltonian with the so-called bearded edge. The method we
describe below applies to both types of edge.

The following gauge transformation makes both $p_1$ and $p_2$ real,
\begin{align}
  \label{gauge}
  \notag
  a_n(k_1) &\rightarrow a_n(k_1) \, e^{-i(n-1)k_1/2}\ , \\
  b_n(k_1) &\rightarrow b_n(k_1) \, e^{-ink_1/2}\ ,
\end{align}
by which $v_i \rightarrow v_i \, e^{ik_1/2}$ and $p_1 \rightarrow
p_1 \, e^{ik_1/2}$, \emph{viz.},
\begin{gather}
  \notag v_1 \rightarrow -(t_2 + t_3) \cos(\phi - \frac{k_1}{2}) + i(t_2  -  t_3) \sin(\phi - \frac{k_1}{2}),\\
  \notag v_2 \rightarrow -(t_2 + t_3) \cos(\phi + \frac{k_1}{2}) - i(t_2 - t_3) \sin(\phi + \frac{k_1}{2}),\\
  \label{vp-gauge}
  p_1 \rightarrow -2 \cos \frac{k_1}{2}.
\end{gather}

In eqns.~\ref{kimat} and \ref{rmat}, a twisted boundary condition
\cite{qi-wu-zhang06-tbc} is used:
\begin{gather}
  \label{tbc}
  \tilde v \equiv \rho \,Uv =
  \begin{pmatrix}
    \tilde v_1 & \tilde v_{12}\\
    \tilde v_{21} & \tilde v_2
  \end{pmatrix}.
\end{gather}
$\rho$ is a real number controlling the ``tunnelling strength''
between the two edges, and $U$ is a unitary $2\times 2$ matrix that
describes an ``isospin-dependent'' phase twisting over the boundary.
For an open boundary, $\rho \rightarrow 0$. For periodic boundary conditions,
$\rho = 1$ with $U = \mathbb{I}$ (without the gauge transformation of eqn.~\ref{gauge})
or $U = e^{-iNk_1/2} \, \mathbb{I}$ (with eqn.~\ref{gauge}), where $\mathbb{I}$ is the
$2\times 2$ identity matrix.  Introducing
twisted boundary condition may seem to overcomplicate the situation,
but as we shall see it allows us to make progress toward an analytical solution.

The eigenvalue problem can now be written as
\begin{gather}
  \notag
  K^{(1)} |\psi_A\rangle + R\, |\psi_B\rangle = \varepsilon\, |\psi_A\rangle\ ,\\
  \label{kr-eig}
  K^{(2)} |\psi_B\rangle + R\yd\, |\psi_A\rangle = \varepsilon\,
  |\psi_B\rangle\ ,
\end{gather}
where $|\psi_A\rangle$ and $|\psi_B\rangle$ are the ``wavefunctions''
of the $A$ and $B$ sublattices, both of which are $N$-dimensional column vectors.

\begin{figure*}[t!]
  \subfigure[\;{\it Ansatz\/} vs. open boundary spectrum]{
    \label{ansatz-erg}\includegraphics[width=0.48\textwidth]{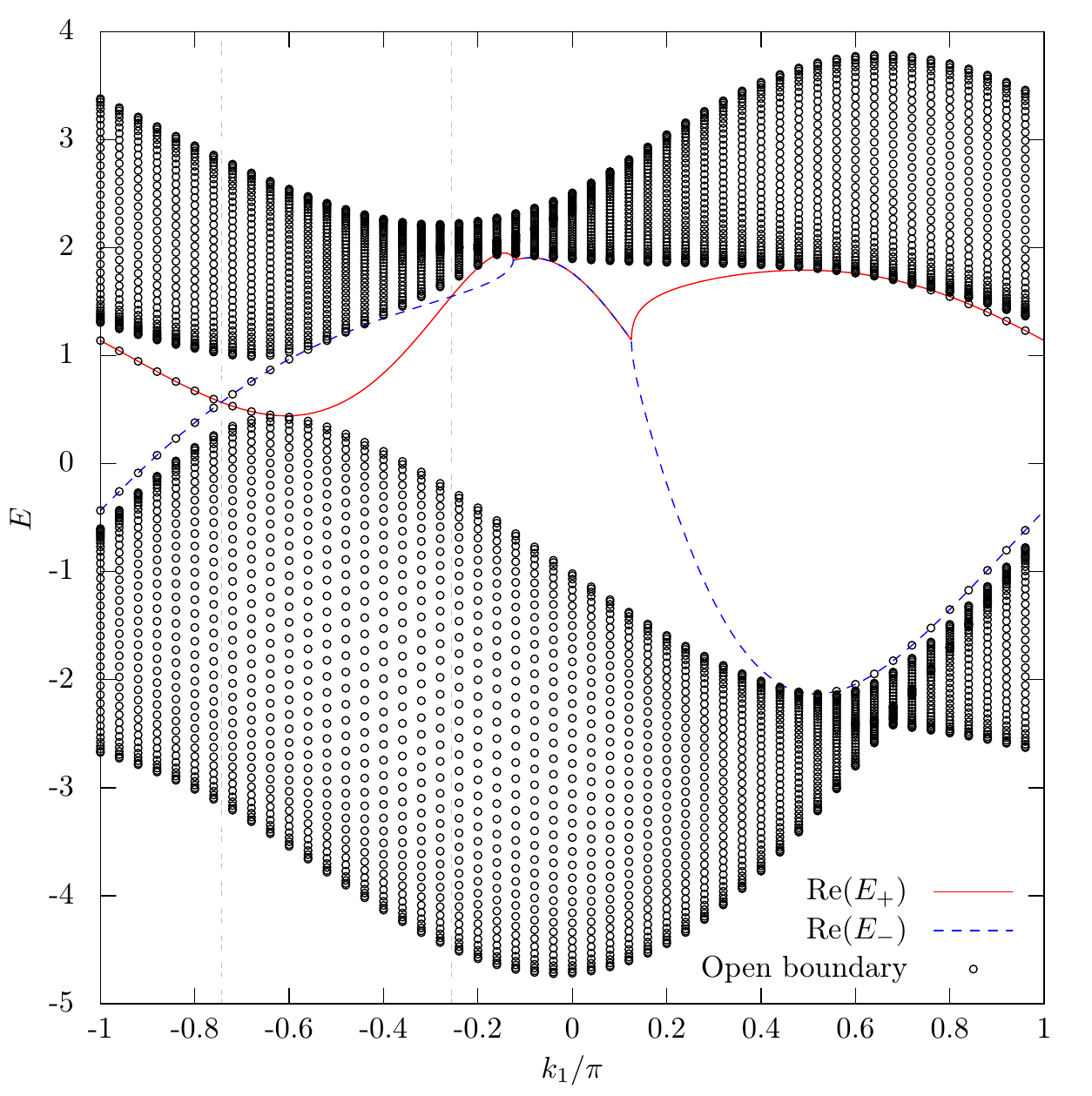}
  }\subfigure[\;auxiliary quantities]{
    \label{ansatz-par}\includegraphics[width=0.48\textwidth]{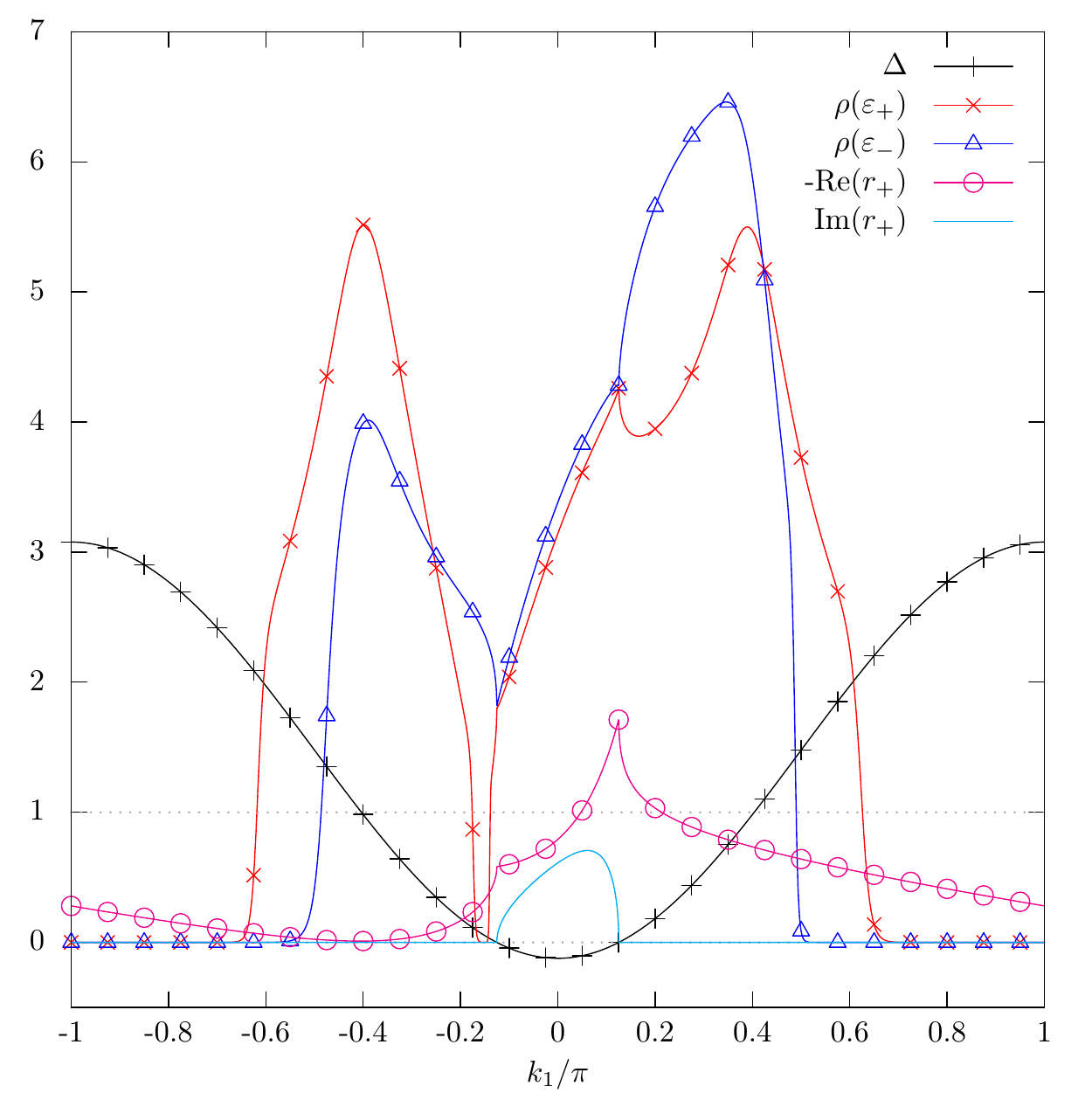}
  }\caption{(Color online) Zigzag edge modes for $[t_1, t_2, t_3] =
    [0.3,0.4,0.5]$, $m = 1.4$, $\phi = 0.3\pi$, $N = 40$. (a)
    Circular dots are obtained from diagonalizing an open boundary
    Hamiltonian, eqn.~\ref{h-tbc} with $\rho =0$. Colored curves are
    {\it Ansatz\/} solutions, eqn.~\ref{e-r}. Gray vertical lines mark the
    edge level crossing points given by eqn.~\ref{kc}. (b) Auxiliary
    quantities. $\Delta$ is the discriminant in
    eqn.~\ref{discriminant}, only $\Delta > 0$ yields real solutions
    of edge energy $\varepsilon_{\pm}$ and is physical. $r_{+}$: the
    ratio of eqn.~\ref{ratio} corresponding to the $\varepsilon_{+}$
    branch. It is real if $\Delta > 0$. $r_{-}$ can be obtained from
    eqn.~\ref{inversion-lr} and is not plotted here.
    $\rho(\varepsilon_{\pm})$: the inter-edge tunnelling strength as
    defined in eqn.~\ref{h-tbc}. These are solved in a self-consist
    fashion in Appendix \ref{zzwf}. $\rho \rightarrow 0$ for open
    boundary, $\rho > 1$ for ``enhanced-tunnelling'' boundary. The
    transition $\rho = 1$ corresponds to bulk modes.}
  \label{ansatz}
\end{figure*}

\subsubsection{Edge state {\it Ansatz\/} and energy}
We look for solutions of the form $|\psi_A\rangle = |\psi\rangle$ and
$|\psi_B\rangle = \lambda |\psi\rangle$. In terms of the
direct-product decomposition discussed earlier, $|\psi\rangle$ is the
real space part and $\left(
  \begin{smallmatrix}
    1\\ \lambda
  \end{smallmatrix}
\right)$ is the internal space part.
Eqn.~\ref{kr-eig} now becomes
\begin{gather}
  \label{edge-eig-eqn-k1}
  (K^{(1)} + \lambda R - \varepsilon) |\psi\rangle = 0\ ,\\
  \label{edge-eig-eqn-k2}
  (\lambda K^{(2)} + R\yd - \lambda\varepsilon) |\psi\rangle =
  0\ .
\end{gather}
A sufficient condition for both equations to be satisfied is that the
coefficient matrices are proportional element by element,
\begin{gather}
  \label{ratio}
  {h_1 + \lambda p_1 - \varepsilon\over h_2 + \lambda^{-1}p_1 -
    \varepsilon}
  = {v_1\over v_2 + \lambda^{-1}p_2}
  = {v_1^{*} + \lambda p_2\over v_2^{*}} \equiv r\ .
\end{gather}
This gives, at each value of $k_1$, two equations (the ratio $r$ itself being yet
undetermined) for the two unknowns $\lambda$ and $\varepsilon$.  The solutions are
\begin{gather}
  \label{lpm}
  \lambda_{\pm} = {v_1v_2^{*} - v_2 v_1^{*} - p_2^2 \pm \sqrt{d^2 - 4|v_1v_2|^2}\over 2 p_2 v_2}\ ,\\
  \label{rpm}
  r_{\pm} \equiv r(\lambda_{\pm}) = {d \pm \sqrt{d^2 - 4|v_1v_2|^2}\over 2|v_2|^2}\ ,\\
  \label{e-r}
  \varepsilon_{\pm} = {r_{\pm}(p_2 h_2 - 2 p_1 \textsf{Re}\,v_2) - (p_2h_1 - 2 p_1 \textsf{Re} \, v_1)\over p_2(r_{\pm}-1)}\ ,
\end{gather}
where \textsf{Re} indicates the real part, $\pm$ denotes the branch of solution, and
\begin{gather}
  \label{d}
  d = v_1v_2^{*} + v_2v_1^{*} - p_2^2 \in \mathbb{R}\ .
\end{gather}
See Appendix ~\ref{zzwf} for details regarding derivation.

Eqn.~\ref{rpm} becomes singular when $v_2 = 0$, which could happen for
the zigzag edge if $t_2 = t_3$ and $k_1 = \pi - 2\phi$. For this
particular parameter set, one can readily verify, by Taylor
expansion in $v_2$, that
\begin{gather}
  \notag
  \lambda_{+} = v_1\ , \quad r_{+} = -v_1^2\ , \quad
  \varepsilon_{+} = {h_2 v_1^2 + 2 p_1v_1 + h_1\over 1+v_1^2}\ ,\\
  \label{singular}
  \lambda_{-}^{-1} = v_2 \rightarrow 0\ , \quad r_{-}^{-1} = - v_2^2
  \rightarrow 0\ , \quad \varepsilon_{-} = h_2\ .
\end{gather}
These results also hold both for graphene ($m=0$) and for boron nitride ($m
\neq 0$). In both cases, second neighbor hoppings are turned off,
rendering $v_1 = v_2 = 0$. Clearly, $\lambda_{+} = r_{+} =
\lambda_{-}^{-1} = r_{-}^{-1} = 0$, and $\varepsilon_{\pm} = \pm m$.
As will be shown in Appendix \ref{zzwf}, $\rho \rightarrow 0 $ if
$|k_1| > 2\pi/3$, so solutions there correspond to edge modes with
open boundary.

A natural question arises regarding the reality of the energy
eqn.~\ref{e-r}. As long as we can find wavefunctions complying with
the {\it Ansatz\/} $|\psi_B\rangle = \lambda \, |\psi_A\rangle$,
$\varepsilon_{\pm}$ will be eigenvalues of a Hermitian matrix, and
hence real. This implies that $r_{\pm}$ are also real
({\it cf.\/}~eqn.~\ref{r-e}). Thus by eqn.~\ref{rpm}, our {\it Ansatz\/} yields real
solutions, with \emph{some} choice of $\rho\,U$, provided the discriminant satisfies
\begin{gather}
  \label{discriminant}
  \Delta = d^2 - 4|v_1v_2|^2 \ge 0\ .
\end{gather}
Note that this condition is valid for all $k_1$.

Although real solutions exist for all $k_1$ with $\Delta(k_1) \ge 0$,
they do not necessarily correspond to open boundaries.  Normally,
wavefunctions are solved after fixing boundary conditions ($\rho$ and
$U$), but here, we enforced a particular form of solution, which will
\emph{not} be consistent with arbitrary $\rho\,U$. Instead, the matrix
$\rho\,U$ is to be determined self-consistently from the {\it Ansatz\/}, and
is in general $k_1$-dependent. This is discussed in detail in Appendix
\ref{zzwf}. For now, we just note that only when $\rho \rightarrow 0$
will the solution be valid for open boundary. Clearly, as $\rho$
varies with $k_1$, the transition from an open boundary solution to
that of an ``enhanced tunnelling'' boundary will happen when $\rho =
1$, at which point the edge solution merges into the bulk
($|\psi\rangle$ becomes extended instead of localized).

Fig.~\ref{ansatz} shows the {\it Ansatz\/} solutions (colored curves)
comparing with the open boundary spectrum (circular dots) in
\ref{ansatz-erg}, and auxiliary variables $\Delta$,
$r(\varepsilon_{+})$ and $\rho(\varepsilon_{\pm})$ in
\ref{ansatz-par}. Parameters are chosen to exhibit most of the
possible scenarios, $[t_1, t_2, t_3] = [0.3,0.4,0.5]$, $m = 1.4$,
$\phi = 0.3\pi$, $N = 40$: \nl (1) $\Delta < 0$ for $|k_1| \lesssim
0.125$. In this region the {\it Ansatz\/} does not yield real energy
solutions. \nl (2) For $|k_1| \gtrsim 0.125$, the {\it Ansatz\/} yields
physical solutions. One then computes $\rho$ self-consistently; $\rho
\ll 1$ means open boundary, whereas $\rho > 1$ means
enhanced-tunneling boundary. The transition happens when the open
boundary edge modes merge with the bulk bands. The $\varepsilon_{+}$
branch (red curve) has open boundary for $k_1 \in [-\pi,-0.62\pi] \cup
[-0.17\pi,-0.14\pi] \cup [0.62\pi, \pi]$, while the $\varepsilon_{-}$
branch (blue curve) has open boundary for $k_1 \in [-\pi, -0.48\pi]
\cup [0.49\pi, \pi]$. In these regions, the {\it Ansatz\/} solutions overlap
with the open-boundary numerics (filled circles). Note that the
$\varepsilon_{+}$ branch briefly becomes open boundary in $[-0.17\pi,
-0.14\pi]$. Without the {\it Ansatz\/} solution, one would have taken it to be
part of the bulk spectrum. \nl (3) Within the physical regime ($\Delta
> 0$), the two branches cross twice, marked by the two vertical gray
lines, one with $\rho(\varepsilon_{\pm}) \rightarrow 0$ and the other
with $\rho(\varepsilon_{\pm}) > 1$. In both cases
$\rho(\varepsilon_{+}) = \rho(\varepsilon_{-})$. These two edge
crossing $k_1$ points are described by eqn.~\ref{kc} which will be
discussed in the next section. While only the one with $\rho
\rightarrow 0$ is relevant for the open-boundary edge spectrum, we
shall see in the next section that both have geometrical significance
and will be reflected in the entanglement spectrum and Berry phase
flow.

\begin{figure*}[t!]
  \centering
  \subfigure[\;Periodic-boundary entanglement occupancy]{
    \label{mlh-ent-f}\includegraphics[width=0.32\textwidth]{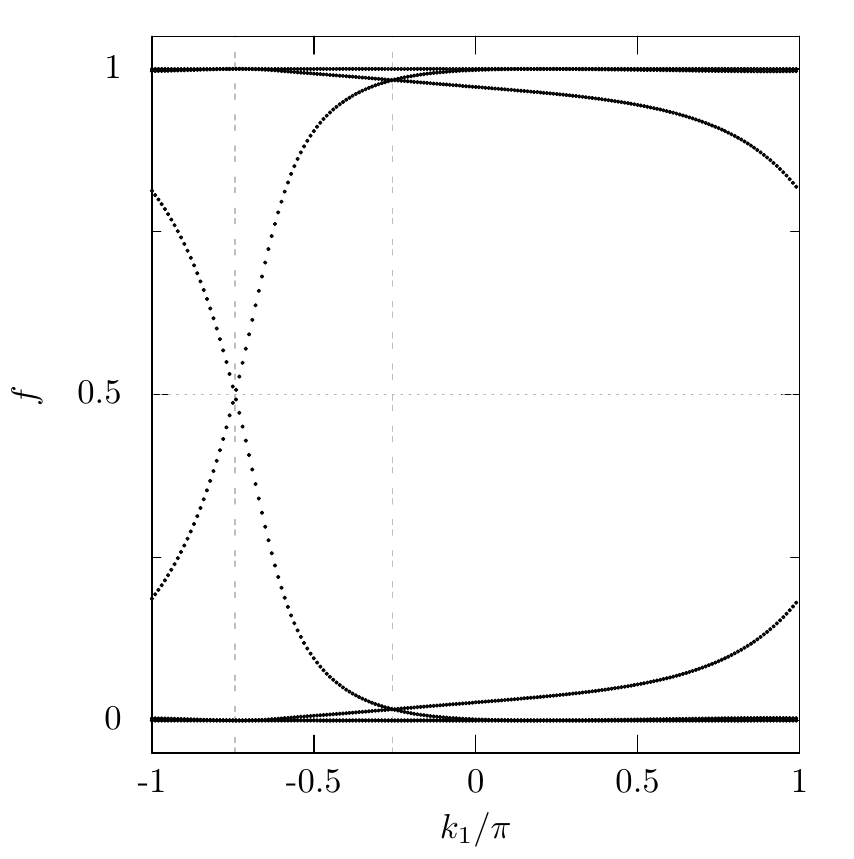}
  }\subfigure[\;Periodic-boundary entanglement quasi-energy]{
    \label{mlh-ent-g}\includegraphics[width=0.32\textwidth]{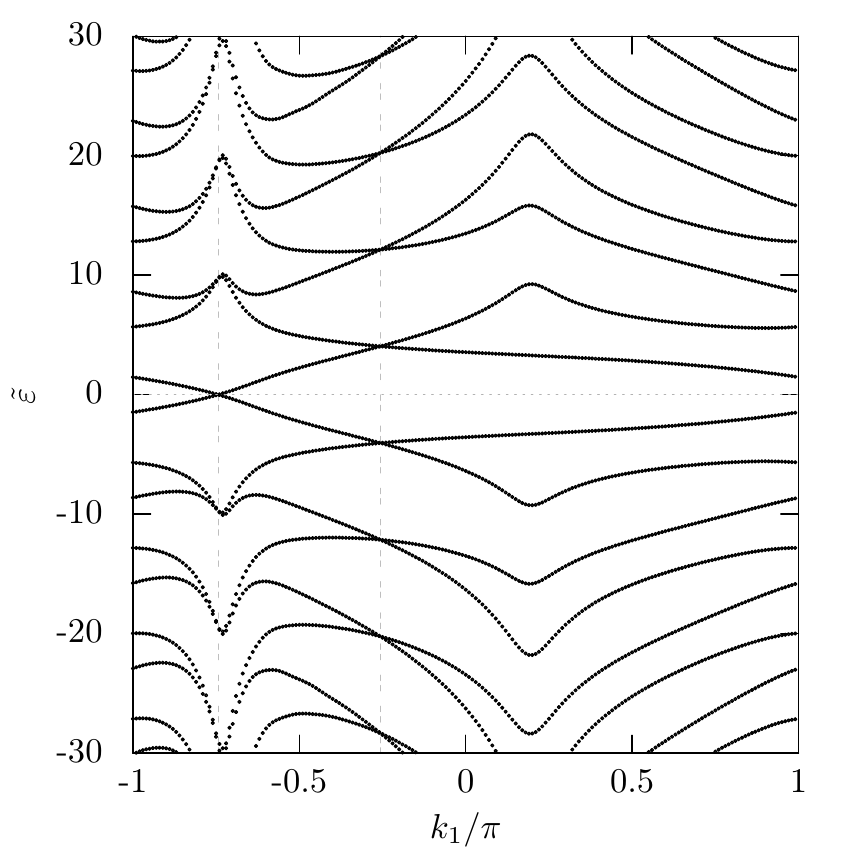}
  }\subfigure[\;Deviation of Wannier centers from unit cell boundary]{
    \label{mlh-berry-phase}\includegraphics[width=0.32\textwidth]{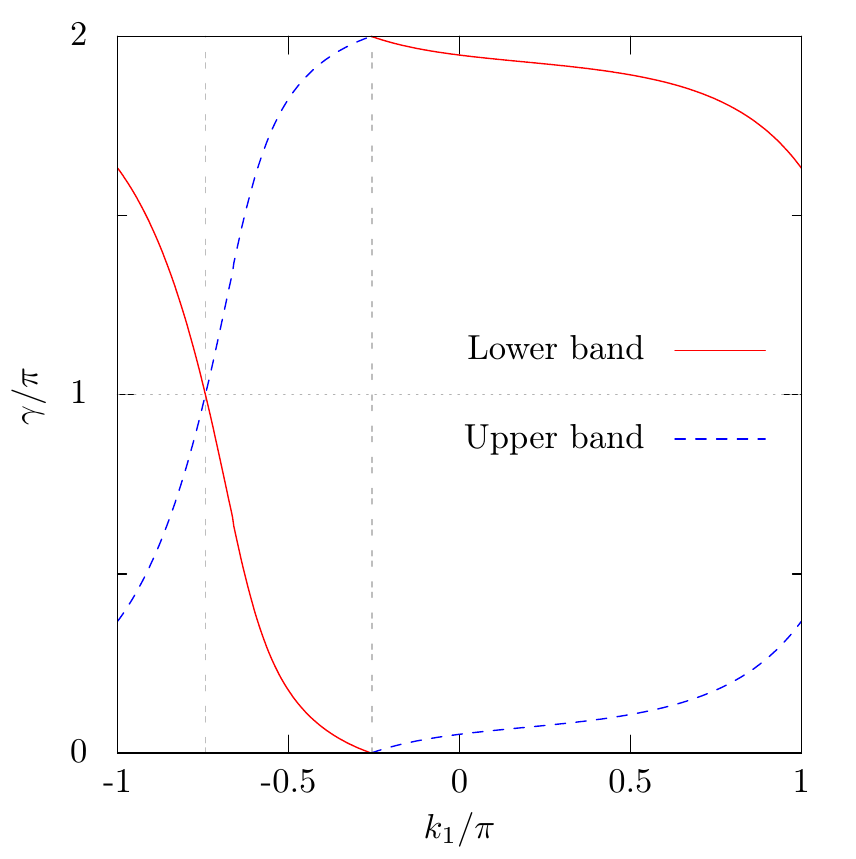}
  } 
  \caption{(Color online) Coincidence of level crossings in
    entanglement and Berry phase flows. (a): Entanglement occupancy.
    (b): Entanglement quasienergy. (c): Deviations of Wannier centers
    from their corresponding unit cell boundary; mathematically this
    is the Berry phase; plotted for both the occupied and unoccupied
    bands. Periodic boundary conditions are used. Parameters used here
    are the same as in Fig.~\ref{ansatz}: $[t_1, t_2, t_3] =
    [0.3,0.4,0.5]$, $m = 1.4$, $\phi = 0.3\pi$. Entanglement spectra
    are computed for the lower half system with $M = N/2 = 20$ unit
    cells in the $\vec a_2$ direction, where $N=40$ is the number of
    unit cells of the full system, and Fermi energy is placed in the
    gap, $\EF = 0.6$. Berry phases are computed using $100$ steps in
    the $k_2$ integration. Notice the similarity between the
    entanglement occupancy and the Berry phases. This is due to the
    former being a coarse-grained version of the latter in the sense
    of Ref.~\onlinecite{ha11-hofstadter}. Notice that in all three
    plots, level crossings occur at the same $k_{\rm c}$ points given by
    eqn.~\ref{kc}. Although the open boundary edge modes only cross at
    one of the two $k_{\rm c}$, the entanglement spectra and the Berry
    phases cross at both $k_{\rm c}$, but with different $f$, $\qe$ and
    $\gamma$ values. $f=\half$, $\qe = 0$ and $\gamma = \pi$ at the
    open-boundary edge crossing point. At the other $k_{\rm c}$ point,
    $\gamma = 0$. Notice also that in the quasienergy plot, level
    crossings are not restricted to the central two levels but extend
    all the way to big $|\gamma|$ values, and are pinned at the same
    $k_{\rm c}$ values.}
  \label{mlh-crossings}
\end{figure*}

\subsection{Topological signatures in edge, entanglement and Wannier spectra }
A gapless edge spectral flow is one of the most conspicuous real-space
manifestation of nontrivial topology of band insulators. But as
discussed in the introduction, it sometimes fails to reveal every
topological difference a band insulator can have from its atomic limit
\cite{hughes-prodan-bernevig11-inversion}, which the entanglement
spectrum can capture. In the non-interacting fermion case, the
entanglement occupancy spectrum is the eigenvalues of a submatrix $G$
of the one-body ground state projector $\g$ \cite{cheong04,peschel03},
with the dimension of $G$ given by the entanglement cut. The
eigenvalues of $\g$ itself are just $0$s and $1$s, but for a system
with nontrivial topology, the eigenvalues of $G$ exhibit a spectral
flow from $0$ to $1$, as a function of the momentum $\kperp$ along the
entanglement cut. The reason why the entanglement cut would induce
such a flow can be understood intuitively in terms of the Wannier
states. One can always recombine the Bloch states that constitute $\g$
(assuming the system is periodic) into a set of spatially localized
Wannier states. If a Wannier state resides in either half of the
partition, it has almost perfect projection onto that half,
and the corresponding entanglement occupancy is very near $0$ or $1$.
If on the other hand the entanglement cut passes right through a
Wannier state (the exact meaning of which we shall discuss in
\S\ref{ent-half-occu}), then it has significant projection onto both
halves, correspondingly the entanglement occupancy is near $\half$. In
the topological phase, the Wannier states themselves flow with respect
to $\kperp$ \cite{Qi11-wannier, soluyanov-vanderbilt11-z2-wannier,
  Yu11-z2-wannier}, thus when one such state passes through the
entanglement cut, a corresponding entanglement flow arises. One can
see that the position of the Wannier states, or the Wannier spectrum,
is closely related to the entanglement spectrum, and that a nontrivial
topology underlies the $f=\half$ entanglement occupancy mode, and a
half-odd-integer Wannier center. It is thus interesting to note that
in the Haldane model, the edge crossing point, the half occupancy mode
and the half-odd-integer Wannier center all coincide at the same $k_1$
point. See Fig.~\ref{mlh-crossings}. In this section we shall study
the reason underlying this coincidence. We mention that there are
several other works \cite{thonhauser06, coh09} studying the charge
polarization of the Haldane model, from the Wannier function
perspective.

\subsubsection{Edge modes crossing points}
\label{ecp}
With the edge solution in \S\ref{mlh-zzedge}, we can identify the
exact location of these crossing points. The condition for
$\varepsilon_{+} = \varepsilon_{-}$ is that
\begin{gather}
  \label{ec}
  p_2(h_1 - h_2) + 2 p_1 \textsf{Re}(v_2 - v_1) = 0\;,
\end{gather}
see Appendix \ref{zzwf} for derivation. For the zigzag edge, this
implies
\begin{gather}
  \label{ec-zz}
  m + 2\sin\phi \sin k_{\rm c}(t_1 + t_2 + t_3) = 0
\end{gather}
whence
\begin{gather}
  \label{kc}
  k_{\rm c} =
  \begin{cases}
    -\sin^{-1} \dfrac{m}{2(t_1 + t_2 + t_3)\sin\phi}\\
    \quad & \text{ or }\\
    \pi + \sin^{-1} \dfrac{m}{2(t_1 + t_2 + t_3)\sin\phi}
  \end{cases}
\end{gather}
Here $k_{\rm c}$ denotes the values of $k_1$ where the two edge modes are energetically degenerate.
The bearded edge solution is obtained by switching the $1$ and $2$ suffixes in
$p_i, h_i$ and $v_i$, for which eqn.~\ref{ec} implies
\begin{multline}
  \label{ec-bearded}
  (m + 2 t_1 \sin\phi\sin k_{\rm c}) \cos \frac{k_{\rm c}}{2} \\+ (t_2 +
  t_3)\sin\phi\sin \frac{k_{\rm c}}{2} = 0.
\end{multline}
This could be recast as a cubic equation for $\tan(k_{\rm c}/2)$, but this
does not afford a particularly simple closed form solution.

One can see from Fig.~\ref{ansatz} that only one of the two $k_{\rm c}$ has
$\rho \rightarrow 0$ and corresponds to an open boundary edge mode.
The other one is of an enhanced-tunnelling boundary; in some parameter
settings it even lies in the region $\Delta < 0$ where the {\it Ansatz\/}
solution is complex. However, the enhanced-tunnelling $k_{\rm c}$ is still
special in the entanglement and Wannier center flows, as can be seen
in Fig.~\ref{mlh-crossings}. What is the significance about these edge
crossing points? Recall that the bulk Hamiltonian $H(k_1, k_2)$ maps
each $\vec k$ to a $\vec B(\vec k)$ vector. Fixing $k_1$ while varying
$k_2$ will drive the $\vec B$ vector along a closed curve in $3$D. It
turns out that at both $k_{\rm c}$, this curve lies \emph{on a plane passing
  through the origin}. To see this, we note that at the edge crossing
point, the $2\times 2$ blocks $h(k_1)$ and $v(k_1)$ are related via
\begin{gather}
  h = \varepsilon_{\rm c} \, \mathbb{I} + {p_1\over p_2} \left(v +
    v\yd\right)
\end{gather}
where $\varepsilon_{\rm c} = \varepsilon_{+}(k_{\rm c}) = \varepsilon_{-}(k_{\rm c})$
(cf.~eqn.~\ref{ece}). The bulk Hamiltonian at the edge crossing points
is then
\begin{align}
  \notag
  H(k_{\rm c}, k_2) &= h + v \, e^{i k_2} + v\yd \, e^{-ik_2}\\
  &= \left[
    v \left( {p_1\over p_2} + e^{ik_2} \right) + {\rm H.c.}
  \right] + \varepsilon_{\rm c} \, \mathbb{I},
\end{align}
and the corresponding $\vec B(k_{\rm c}, k_2)$ is
\begin{gather}
  \notag
  B_z = \left( {p_1\over p_2} + \cos k_2 \right) \textsf{Re}\,(v_1 - v_2) -
  \sin k_2 \,\textsf{Im}\,(v_1 - v_2) \;,\\
  \label{bkc}
  B_x = p_1 + p_2 \cos k_2\;, \quad B_y = p_2\sin k_2\;,
\end{gather}
note in particular that all components are independent of $m$, which
is the term that breaks the inversion of the two sublattices. It is then easy to check that
\begin{gather}
  \vec B(k_{\rm c}, k_2) = B_x \left(\hat x - \textsf{Re}(q)\>\hat z\right) +
  B_y\left(\hat y - \textsf{Im}(q) \>\hat z\right)
\end{gather}
where
\begin{gather}
  q = {v_2 - v_1\over p_2}
\end{gather}
is independent of $k_2$. The path of $\vec B(k_{\rm c}, k_2)$ is thus
coplanar and normal to the vector
\begin{gather}
  \label{mlh-ec-norm}
  \vec n \equiv \textsf{Re}(q)\> \hat x +\textsf{Im}(q)\>\hat y + \hat z.
\end{gather}
An unrestricted $(B_x, B_y)$ pair can describe any point on the plane.  Clearly the origin itself
is on the plane.  The actual path of $\vec B(k_{\rm c}, k_2)$ is restricted to those allowed by
eqn.~\ref{bkc}.

An interesting observation is that in graphene, the bulk $2\times 2$
Hamiltonian is always off-diagonal, thus it is coplanar at any $k_1$
value. The origin is inside the path of $\vec B$ for $|k_1| > 2\pi/3$,
and outside otherwise, hence the well known result that its two zigzag
edge modes are degenerate at $\varepsilon = 0$ for $|k_1| > 2\pi/3$.
For the bearded edge, the degenerate edge modes connect the two Dirac
points in the other way, namely with $|k_1| < 2\pi/3$.

\subsubsection{Integer and half-odd-integer Wannier centers}
For a general one-dimensional periodic system, or higher-dimensional system with
$\kperp$ specified, the Wannier centers can be defined as the
nonvanishing eigenvalues of the band-projected real-space operator
\begin{gather}
  \label{grg}
  \g(\kperp) \,  R  \> \g(\kperp)
\end{gather}
where $\g(\kperp)$ is the filled band projector (or sum of projectors for multiple
bands) at $\kperp$, $R = \hat Y \!\otimes\! \mathbb{I}$ in which $\hat Y =
\textsf{diag}(1,2, \ldots, N)$ measures real space (\emph{i.e.}, unit
cell) coordinates, $N$ being the number of unit cells in the
longitudinal direction, and $\mathbb{I}$ is the $q\times q$ unity
acting on the $q$-dimensional internal space. The corresponding
eigenstates are defined as the Wannier states
\cite{kivelson82-wannier}. The monolayer Haldane model has only one
band occupied at half filling. When $\g$ contains only one band, the
eigenvalues of eqn.~\ref{grg} are \cite{Qi11-wannier}
\begin{gather}
  \label{wannier-1b}
  \lambda_I(\kperp) = {\gamma(\kperp)\over 2\pi} + I\quad , \quad I =
  1, 2, \ldots N
\end{gather}
where $I$ labels the unit cell, and $\gamma(\kperp)$ is the Berry
phase of the band,
\begin{gather}
  \gamma(\kperp) = \int\limits_0^{2\pi} \!\! d\kpara A(\kpara)\quad, \quad
  A(\kpara) \equiv i\, \langle \psi | \, \partial_{\kpara} \, | \psi\rangle\;.
\end{gather}
The $\kperp$ dependence of $A$ is suppressed. $A$ is the $\kpara$
component of the Berry connection vector at fixed $\kperp$. Thus the
offset of the Wannier centers from the unit cell boundaries are
uniform and are given by the Berry phase\cite{zak89-berry-phase}.

In the Haldane model, $\kperp = k_1$ and $\kpara = k_2$. The lower
band projector is
\begin{gather}
  \g(k_1) = \sum_{k_2} |\Psi(\vec k) \rangle \langle \Psi(\vec k)| \;,\\
  \label{bloch}
  |\Psi(\vec k)\rangle \equiv |k_2\rangle \otimes |\psi(k_1,
  k_2)\rangle\;,
\end{gather}
where the internal part $|\psi(k_1, k_2)\rangle$ is the lower band
eigenstate of the bulk Hamiltonian eqn.~\ref{hbulk}, and the
real-space part $|k_2\rangle$ is the Bloch wave in the $\vec a_2$
direction, $\langle y | k_2\rangle = e^{ik_2 y}/\sqrt{N}$. The upper
band projector is similarly defined. We plot the Berry phase in
Fig.~\ref{mlh-berry-phase}, where the red and blue curve correspond to
the lower and upper band, respectively. We will show below that
coplanarity of $\vec B$ at $k_{\rm c}$ fixes the Berry phases there to be
either $\pi$ or $0$, depending on whether or not the origin lies
inside the path of $\vec B$. In fact, this holds for any two-band
model with coplanar $k$ points. Recall $H(\vec k) = \omega(\vec k)
\mathbb{I} + \vec B(\vec k) \cdot \vec \sigma$. Let $\left(B(\vec k),
  \vartheta(\vec k), \varphi(\vec k)\right)$ be the spherical
coordinate of $\vec B(\vec k)$, then the lower band eigenstate is
\begin{gather}
  \label{mlh-psi}
  |\psi\rangle =
  \begin{pmatrix}
    -\sin \frac{\vartheta}{2}\\
    e^{i\varphi} \cos \frac{\vartheta}{2}
  \end{pmatrix}\;,
\end{gather}
whence
\begin{gather}
  A(k_2) = - \frac{1}{2}(1 + \cos\vartheta) {\partial
    \varphi\over \partial k_2}
\end{gather}
and
\begin{gather}
  \label{mlh-gamma}
  \gamma(k_1) = - \half\!\!\!\!\int\limits_{\varphi(0)}^{\varphi(2\pi)}\!\!\!\!\!
  d\varphi \,(1 + \cos \vartheta)\;,
\end{gather}
where $\varphi(0) = \varphi(k_1, k_2 = 0)$ and $\varphi(2\pi) =
\varphi(k_1, k_2 = 2\pi)$. At the edge crossing $k_1$, $\hat B$ is
comprised of points on a great circle (since the origin is on the
plane), so one can always rotate the internal space to a frame where
the plane normal $\vec n$ (eqn.~\ref{mlh-ec-norm}) coincides with the
$\hat z$ axis, then $\cos \vartheta = 0$ everywhere on the coplanar
path which is now lying in the new $xy$ plane. The Berry phases at
coplanar $k_1$ points are simply $2\pi$ times half the (negative)
winding number $w$ of the path of $\vec B$,
\begin{gather}
  \label{winding}
  \gamma(k_{\rm c}) = - w\pi\quad,\quad w \equiv
  {\varphi(2\pi) - \varphi(0)\over 2\pi}\;.
\end{gather}
Note that this rotation amounts to applying the following unitary
transformation to the band projector $\g$ in eqn.~\ref{grg},
\begin{gather}
  \g \rightarrow \tilde\g = U \g\, U\yd \quad , \quad U =
  \mathbb{I} \otimes D(\hat n)
\end{gather}
where $D(\hat n)$ is the aforementioned internal space rotation that
depends only on $\hat n$. Since $UR\,U\yd = R$, both $\g R\, \g$ and
$\tilde \g R \, \tilde \g$ have the same spectrum, so the resulting Wannier
centers and Berry phases are independent of whether $\g$ or $\tilde
\g$ is used.

For the Haldane model, the topological phase has Chern number $|C|=1$,
so the winding number at any $k_1$ is at most $|w| = 1$, hence
$\gamma(k_{\rm c})$ is either $\pm\pi$ (origin inside) or $0$ (origin
outside), which is what we see in Fig.~\ref{mlh-berry-phase}. In the
non-topological phase, $\gamma$ at both $k_{\rm c}$ will be $0$.

\subsubsection{Entanglement Half Occupancy Mode}
\label{ent-half-occu}
As discussed in the introduction, the entanglement spectrum can be
associated with a coarse-graining of the Wannier centers. By
coarse-graining, we mean replacing the real-space operator $R = \hat
Y \otimes 1$ with the projector
\begin{gather}
  P_{M} = f(\hat Y, M) \otimes 1\;,\\
  f(y,M) =
  \begin{cases}
    1 & \text{if }y < M + \frac{1}{2}\\
    0 & \text{if }y > M + \frac{1}{2} \ ,
  \end{cases}
\end{gather}
where $M$ is an integer. In so doing, all Wannier center flows except
those between unit cells $M$ and $M+1$ are suppressed. Since $\g$ and
$P_M$ are both projectors, $\g P_M \g$ and $P_M \g P_M$ share the same
spectrum (cf.~Appendix \ref{pprod}). The latter is nothing but the
restricted correlation matrix whose eigenvalues are the entanglement
occupancy spectrum, and the integer $M$ corresponds to the
entanglement cut. The entanglement eigenstates, $|\tilde \Psi\rangle$,
are projections of eigenstates $|\Psi\rangle$ of $\g P_M \g$ onto
the half space $y \le M$ by $P_M$, {\it i.e.\/} $|\tilde \Psi\rangle = P_M |\Psi\rangle$.

We found that at the edge crossing point $k_{\rm c}$, in the case of odd winding
number, the periodic boundary entanglement occupancy spectrum has two
modes intersecting at $f=\half$ (Fig.~\ref{mlh-ent-f}), which
corresponds to a zero entanglement quasi-energy
(Fig.~\ref{mlh-ent-g}). It is evident from Fig.~\ref{mlh-ent-g} that
the entanglement quasi-energy spectrum has a particle-hole symmetry
for all $k_1$. This is a consequence of the periodic boundary
condition: there are two independent flows in Figs.~\ref{mlh-ent-f}
and \ref{mlh-ent-g} -- one upward, one downward -- because by using
periodic boundary condition, an entanglement cut creates two new
boundaries in each of the half systems. These two flows intersect at
both $k_{\rm c}$ points. Had we started with a $\g$ for an open boundary
system (in the $\vec a_2$ direction), then the entanglement cut would
only create one new boundary to each of the half systems, and as a
result, in each half system, only one of the two flows, which
corresponds to the edge created by the entanglement cut, would survive. In
that case, there would be no particle-hole symmetry in the entanglement
quasi-energy spectrum for arbitrary $k_1$, but only at the edge crossing
$k_{\rm c}$ points, where the entanglement spectrum still exhibits particle-hole
symmetry, since switching the $\vec a_2$ boundary condition from periodic
to open merely removes the double degeneracy at these points.

In fact, one can argue that in an open boundary system, the $f =\half$
level is a consequence of this survival of entanglement particle-hole
symmetry at the edge crossing point. Assume the number of unit cells
in the full system is $N$ and $N$ is even. In an open boundary system,
there are two edge levels, thus each bulk band has $N-1$ bulk levels.
At the edge crossing point, both edge levels are either below or above
the Fermi energy, so the rank of $\g$ is $N \pm 1$ which is always
odd. We may always pick the one half of the bipartite whose size $M
\ge \text{rank}(\g)$, so that $\text{rank}(G) = \text{rank}(\g)$. Thus
with the entanglement particle-hole symmetry at the $k_{\rm c}$ point, there
must be an odd number of entanglement levels fixed at $f = \half$.

With a periodic boundary, the half occupancy modes will appear in the
thermodynamic limit, and for small $N$, there is avoided crossing
between the two flows. However, the degeneracy is exact for arbitrary
even $N$ and $M = N/2$ when $t_2 = t_3$. We prove this in Appendix
\ref{zzent-zm}. Below, we wish to discuss it following the
coarse-graining perspective.

The effect of inserting $R$ and $P_M$ in between two band projectors
$\g$ is to resolve, in real space, the Bloch states that constitute
$\g$. As a result, both the Wannier states and the entanglement states
are spatially localized. In this respect, there exists a family of
such real-space resolvers $R_{\beta}$, parameterized by an ``inverse
temperature'' $\beta$, that smoothly interpolates between the Wannier
and entanglement limits,
\begin{gather}
  \label{fbeta}
  f_{\beta}(\hat Y) \equiv {\tanh\left[\frac{1}{2}\beta\left(\hat
        Y - \frac{1}{2}(N+1)\right)
    \right]\over\tanh\left[\frac{1}{4}\beta(N-1)\right]}\;,\\
  \label{rbeta-eqn}
  R_{\beta} \equiv \frac{1}{2} \! \left[(N-1) f_{\beta}(\hat Y) + N + 1 \right]\otimes \mathbb{I}\;.
\end{gather}
Here, $f_{\beta}$ is obtained by rescaling and shifting the Fermi
distribution. The extremal cases are
\begin{gather}
  \label{rbeta-wannier}
  R_{\beta \rightarrow 0} = \hat Y \;,\\
  \label{rbeta-ent}
  {R_{\beta \rightarrow \infty} - \mathbb{I}\over N-1} = \mathbb{I} - P_{N/2}\;,
\end{gather}
where $\mathbb{I} - P_{N/2}$ is a real space projector onto the $y >
N/2$ half. Thus by varying $\beta$, the spectrum $\{r_a\}$ of $\g
R_{\beta} \,\g$ morphs from the Wannier center spectrum to the
entanglement spectrum, providing a one-to-one mapping between the
Wannier states and the entanglement states.

\begin{figure}[htb!]
  \centering
  \includegraphics[width=0.45\textwidth]{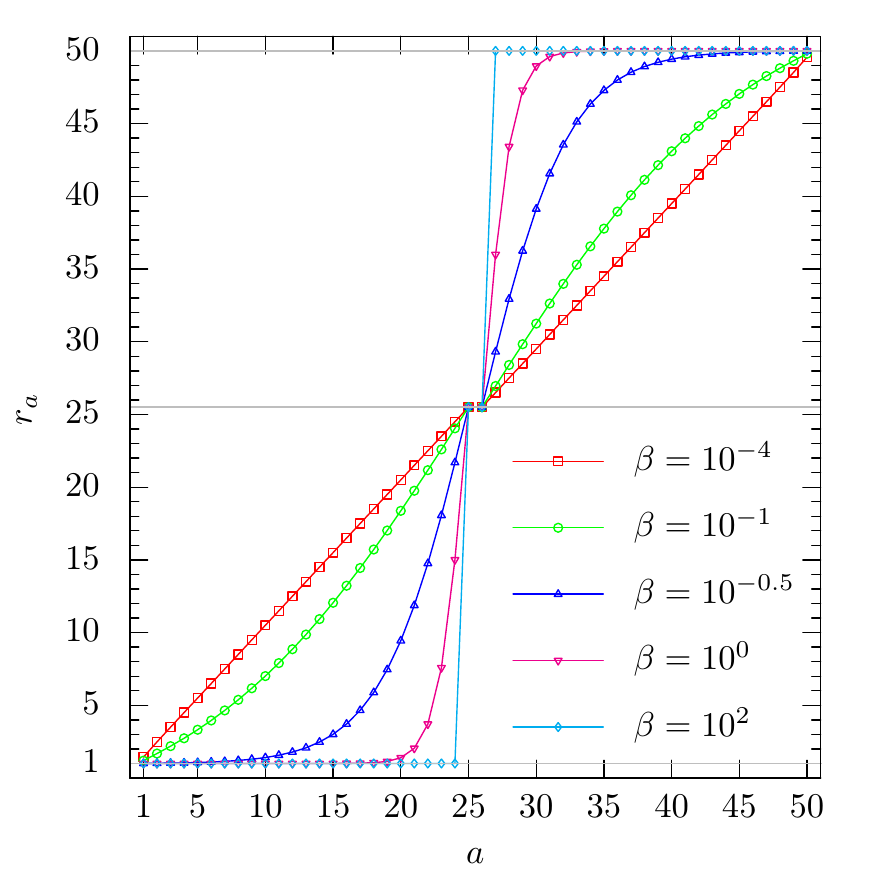}
  \caption{(Color online) Interpolation between Wannier spectrum (red)
    and entanglement spectrum (light blue) at the edge crossing point
    $k_{\rm c}$ with winding number $w = 1$. $N=50$ unit cells in the $\vec
    a_2$ direction, with periodic boundary condition, is used. Other
    parameters are the same as those used in previous figures: $[t_1,
    t_2, t_3] = [0.3,0.4,0.5]$, $m = 1.4$, $\phi = 0.3\pi$. $\{r_a\}$
    is the spectrum of $\g R_{\beta} \,\g$, where $a$ is the level index.
    Different lines correspond to different values of $\beta$. The red
    dots ($\beta \rightarrow 0$) correspond to the limit of Wannier
    center spectrum, therefore the vertical coordinates are pinned at
    half integers (recall the Berry phase is $\pi$). The light blue
    dots ($\beta \rightarrow \infty$) correspond to the entanglement
    spectrum (rescaled and shifted according to eqn.~\ref{rbeta-ent}),
    by which the mid gap levels at $r_a = 25.5$ translate to an
    occupancy of $f=0.5$, $r_a = 1$ to $f=0$ and $r_a = 50$ to $f=1$.
    All curves exhibit particle-hole symmetry (with respect to $r_a =
    25.5$) due to the coplanarity of $\vec B$ at $k_{\rm c}$. Notice the two
    levels pinned at $r_a = 25.5$ for all $\beta$. Had we used open
    boundary condition in obtaining $\g$, there would be only one such
    level.}
  \label{mlh-rbeta}
\end{figure}

\begin{figure}[tb!]
  \centering
  \includegraphics[width=0.49\textwidth]{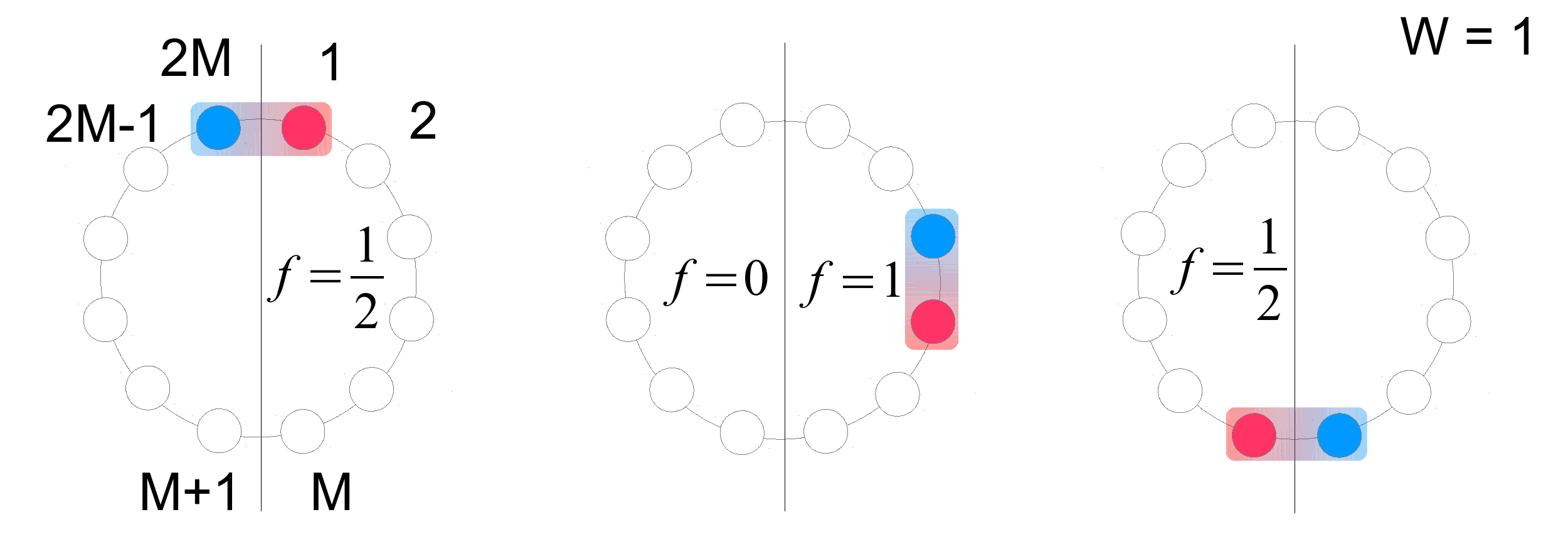}
  \caption{(Color online) Schematics of localization of the two
    sublattices at the edge crossing point with winding number $w =
    1$. $N \equiv 2M$ is the number of unit cells. Each empty dot
    represents a unit cell. Red and blue dots represent the $A$ and
    $B$ sublattice wavefunctions of a Wannier state, respectively. The
    $A$ sublattice is $w = 1$ unit cell ahead of the $B$ sublattice.
    The Wannier state itself is represented as the box around the
    colored dots. The complete set of Wannier states is generated by
    shifting a given Wannier state by integer number of unit cells.
    Since the two sublattices are localized on neighboring unit cells,
    the Wannier center, which is their average spatial location, is a
    half odd-integer. The entanglement eigenstates are qualitatively
    the same as the Wannier states in terms of sublattice
    localization, although quantitatively the wavefunctions do have
    projections onto neighboring Wannier states. If the bipartition
    cut passes through the box (representing the dominant Wannier
    state projection of the entanglement eigenstate), then each half
    has only one sublattice manifesting as a $f=\half$ occupancy mode.
    For periodic systems, there are two such modes, illustrated in the
    left and right panels. Clearly, only the right panel survives as
    entanglement $f=\half$ mode for open boundary systems. The middle
    panel is a Wannier state whose localization does not cross the
    bipartition cut. Qualitatively, it represents an entanglement
    state close to $f=0$ for the left half system and $f=1$ for the
    right half system, cf.~the light blue line in
    Fig.~\ref{mlh-rbeta}.}
  \label{mlh-winding}
\end{figure}

At the coplanar point $k_{\rm c}$ with winding number $w=1$, two $r_a$
levels are pinned at $(N+1)/2$ regardless of the value of $\beta$.
This is shown in Fig.~\ref{mlh-rbeta}. The $\beta$-independence
suggests a unifying physical origin underlying these levels. We have
shown before that, mathematically, the reason for the Wannier spectrum
($\beta \rightarrow 0$) to exhibit such levels is because the Berry
phase of the occupied band, which corresponds to the uniform deviation
of all Wannier centers from the integers labeling their unit cells, is
precisely $\pi$ at the $w=1$ coplanar point. It is instructive to
reflect on the physical implication of this result. For convenience, let us rotate to
the internal basis where the path of $\vec B$ lies in the $xy$ plane,
cf.~discussion leading to eqn.~\ref{winding}. The Berry connection and
its cumulation along the $\vec B$ path are thus
\begin{gather}
  A(k_2) = - {1\over 2} \, {\partial \varphi\over\partial k_2}, \
  \int\limits_0^{k_2}\!\! dk \> A(k) = -\frac{1}{2}\left(\bar\varphi(k_2) + w
    k_2\right)
\end{gather}
where $\bar\varphi(k_2) \equiv \varphi(k_2) - w k_2$ is the
non-winding deviation of $\varphi(k_2)$ from a pure winding term. The
Wannier state corresponding to the $I^{\rm th}$ unit cell is a linear
combination of the Bloch states belonging to the occupied band,
\begin{gather}
  |\Phi_I\rangle = \sum_{k_2} f^{(I)}_{k_2}\, |\Psi(k_2)\rangle \quad ,
  \quad I = 1, 2, \ldots, N
\end{gather}
where $|\Psi(k_2)\rangle$ is the Bloch state defined in
eqn.~\ref{bloch}, and the coefficients $f$ are
\cite{Qi11-wannier}
\begin{gather}
  f^{(I)}_k = \frac{1}{\sqrt{N}} \exp\left(
    -i Ik - \frac{i}{2} \,\bar\varphi(k)\right)\;.
\end{gather}
Recall that in the rotated frame, the internal-space Bloch states of
the occupied band are $ |\psi(k_2)\rangle = \frac{1}{\sqrt{2}}\left(
  \begin{smallmatrix} -1\\e^{i\varphi}
  \end{smallmatrix}
\right)$, thus the Wannier wavefunction of the rotated $A$
and $B$ ``sublattices'' -- which are linear combinations of the
original $A$ and $B$ sublattices in the same unit cell -- are
\begin{gather}
  |\Phi_I^{(A)}\rangle = - \frac{1}{\sqrt{2}}\sum_{k_2} f^{(I)}_{k_2} \, |k_2\rangle = - \frac{1}{\sqrt{2}}
  \hat S\yd(\bar\varphi) \, | I\rangle\;,\\
  |\Phi_I^{(B)}\rangle = \frac{1}{\sqrt{2}}\sum_{k_2} e^{i\varphi(k_2)} f^{(I)}_{k_2} \,
  |k_2\rangle = \frac{1}{\sqrt{2}}\hat S(\bar\varphi) \, | I-w\rangle\;.
\end{gather}
Here, $|I\rangle = e^{-iIk_2}|k_2\rangle/\sqrt{N}$ is the $I^{\rm th}$
real-space basis vector; the unitary \emph{profile} operator $\hat
S(\bar\varphi)$ is
\begin{gather}
  \hat S(\bar\varphi) \equiv \sum_{k_2} |k_2\rangle
  \exp\left(\frac{i}{2}\bar\varphi(k_2)\right)\langle k_2|
\end{gather}
and its effect on $|I\rangle$ is to generate a wavepacket centered at
the $I$, with its profile determined by the details of the fluctuation
$\bar\varphi(k)$. We can then conclude that at the coplanar $k_{\rm c}$
points, the two sublattices of the Wannier states (in the rotated
frame) are localized in \emph{different} unit cells with a separation
given by the winding number $w$ of the path of $\vec B$. The Wannier
centers are spatial averages of the locations of the two sublattice
wavefunctions, $\lambda_I = I - w/2$, thus for odd $w$, they are
half-odd-integers. The double degeneracy as evidenced in
Fig.~\ref{mlh-rbeta} is due to the periodic boundary condition,
because for $I=w=1$, $I-w$ is $N$ instead of zero, thus its Wannier
center is at $(N+1)/2$ which is degenerate with the level of $I =
N/2+1$. This is schematically shown in Fig.~\ref{mlh-winding}. As
$\beta$ deviates from zero, the eigenstates of $\g R_{\beta} \, \g$ leak
into neighboring Wannier states but are still dominated by the $\beta
= 0$ component, thus the physical picture stays the same. When $\beta$
is infinite, the two degenerate Wannier states become the
entanglement half occupancy modes (before projecting onto half
systems) since the entanglement cut assigns the two sublattice
wavepackets into different half systems, leaving each half system with
a ``half electron''.

There is an interesting consequence in entanglement spectrum due to
the $w$-separation between the sublattices. For a general $|w| \neq
1$, it is obvious from Fig.~\ref{mlh-winding} that for each Wannier
state, there are $w$ ways to place an entanglement cut that will
assign the two sublattice wave packets into different halves,
generating a half occupancy mode. Thus in the thermodynamic limit,
there are $2w$ modes of $f = \half$ with a periodic-boundary system, and
$w$ such modes with an open-boundary system. We have confirmed this
with hand-crafted $\varphi(k_2)$ relations, \emph{e.g.}, $\varphi(k_2)
= w\sqrt{2\pi k_2}$. This agrees with the general observation that the
number of entanglement spectral flows is given by the Chern index.

\subsection{Armchair Edge}
The armchair edge is more conveniently described by a new set of
primitive vectors $\vec a_i^{\,\prime}$:
\begin{gather}
  \vec a_1^{\,\prime} = \vec a_1 + \vec a_2\quad , \quad \vec a_2^{\,\prime} = \vec a_2\ ,
\end{gather}
and the associated Bloch phases $k_i' = \vec k \cdot \vec a_i^{\,\prime}$ are
\begin{gather}
  k_1' = k_1 + k_2\quad , \quad k_2' = k_2\ .
\end{gather}
The armchair edge is parallel to $\vec a_1^{\,\prime}$ and the corresponding
open boundary Hamiltonian can be parameterized by $k_1'$. The bulk
Hamiltonian is still eqn.~\ref{omega-b}. When $k_1' = 0$, then $B_y = 0$
and the path of $\vec B$ upon varying $k_2'$ is on the $zx$ plane.
Thus in the topological phase (origin inside the path), one can repeat
the same analysis as done for the zigzag edge and conclude that at
$k_1' = 0$, the Berry phase is $\pi$, and the entanglement occupancy
is $\half$. We have confirmed these numerically. As for the energy edge
modes, while there appear to be no explicit solution, we have
confirmed numerically that they still cross at this coplanar point.

From eqn.~\ref{omega-b}, $B_y / (1+B_x) = \tan(k_1'/2)$, \emph{i.e.},
the projection of $\vec B$ onto the $xy$ plane is a line passing
through the point $(B_x, B_y) = (-1,0)$, so the path of $\vec B$ is
always coplanar for any fixed $k_1'$, but only the one with $k_1' = 0$
has the origin on the same plane. Thus unlike the zigzag edge case,
there exists no other $k_1'$ points whose Berry phase can be related
solely to the winding of a planar $\vec B$ path.

\section{Bilayer Haldane Model}
\begin{figure}[t]
  \centering
  \includegraphics[width=0.4\textwidth]{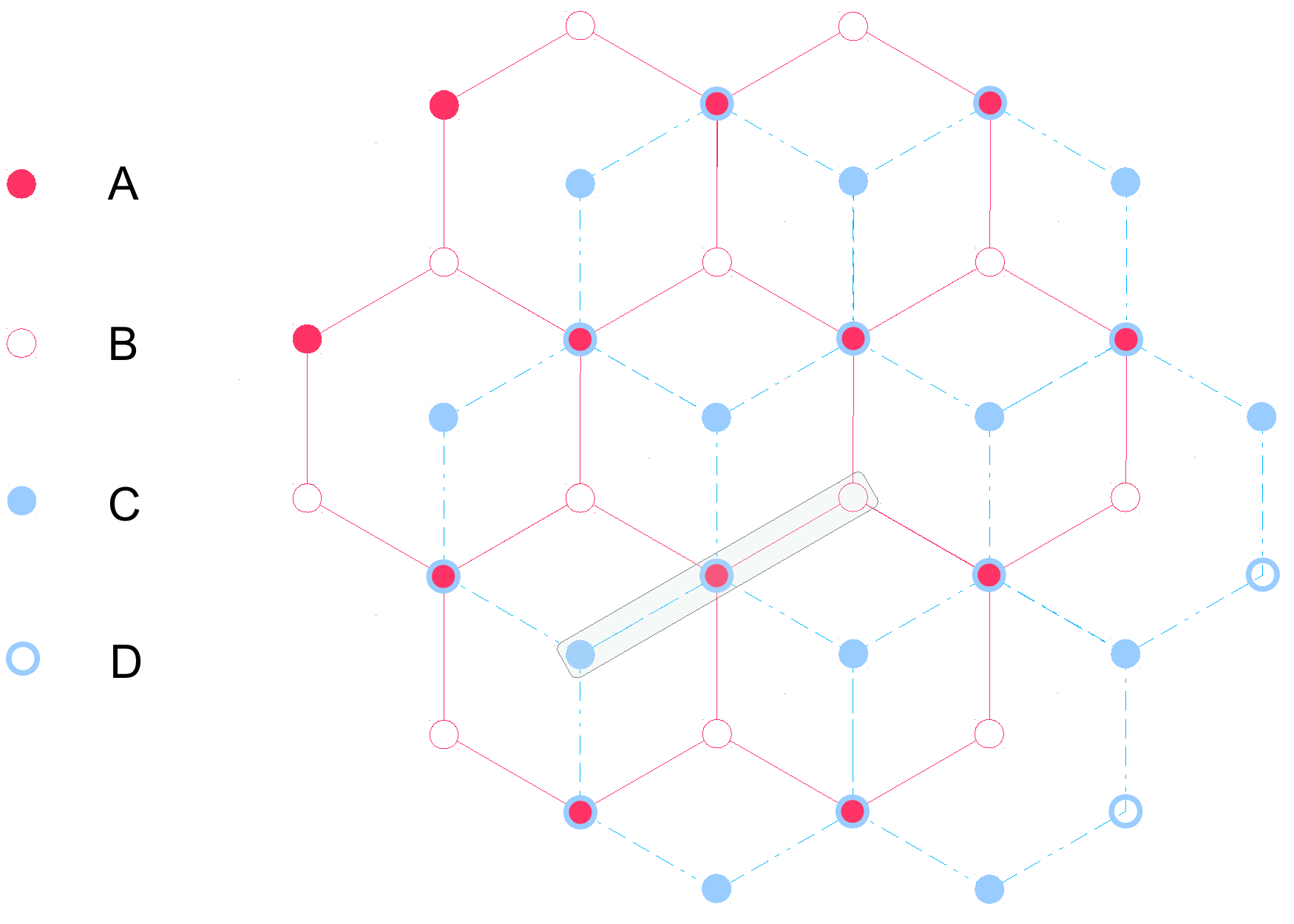}
  \caption{(Color online) Bilayer model. $A$,$B$, $C$ and $D$ label
    sublattice sites, where $A$ and $B$ are on the first layer (solid
    line lattice) and $C$ and $D$ on the second (dashed line lattice).
    Box represents a unit cell. Interlayer hopping is between $A$ and
    the $D$ on top of it.}
  \label{blh}
\end{figure}

\begin{figure*}[t!]
  \subfigure[\;$\phi = 0.3\pi$, $\chi = 0.4\pi$]{
    \label{blh-c2}\includegraphics[width=0.25\textwidth]{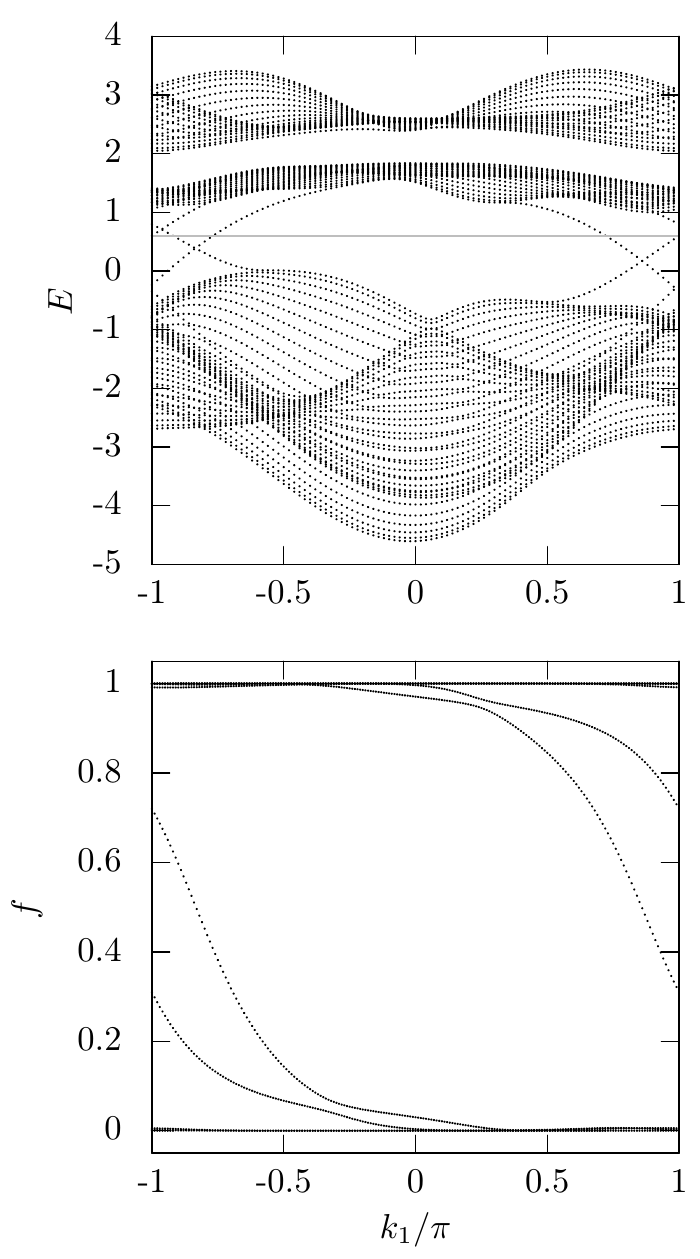}
  }
  \subfigure[\;$\phi = 0.3\pi$, $\chi = 0.05\pi$]{
    \label{blh-c1}\includegraphics[width=0.228\textwidth]{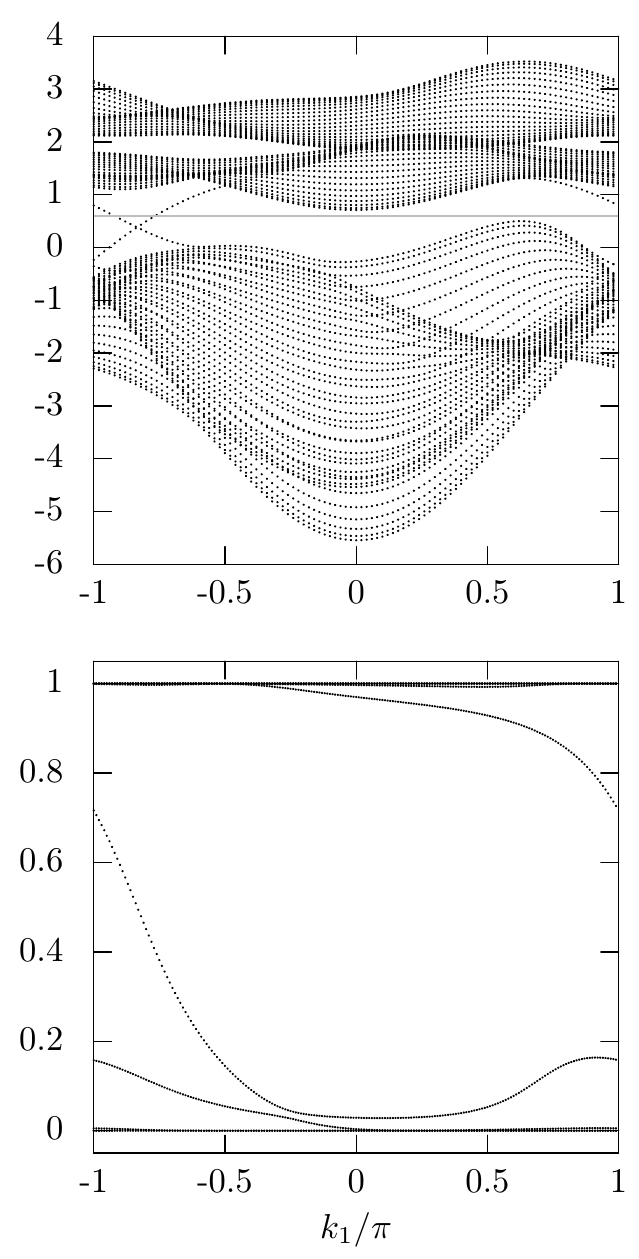}
  }
  \subfigure[\;$\phi = 0.3\pi$, $\chi = -0.4\pi$]{
    \label{blh-c0}\includegraphics[width=0.228\textwidth]{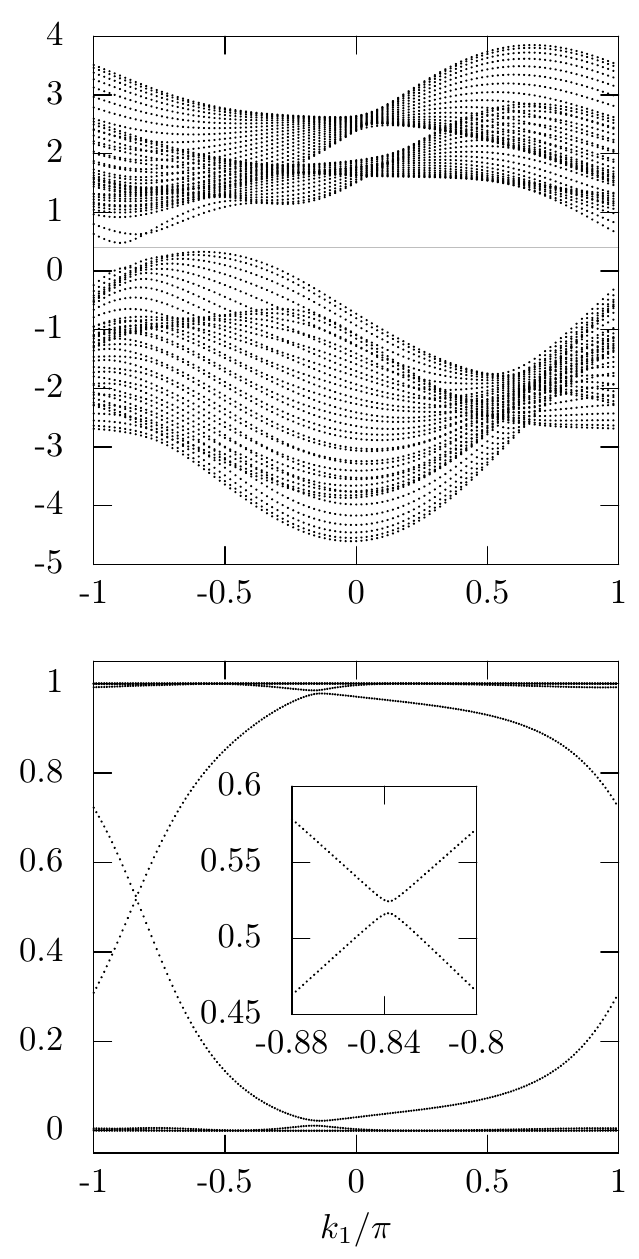}
  }
  \subfigure[\;$\phi = 0.3\pi$, $\chi = -0.3\pi$]{
    \label{blh-c0i}\includegraphics[width=0.228\textwidth]{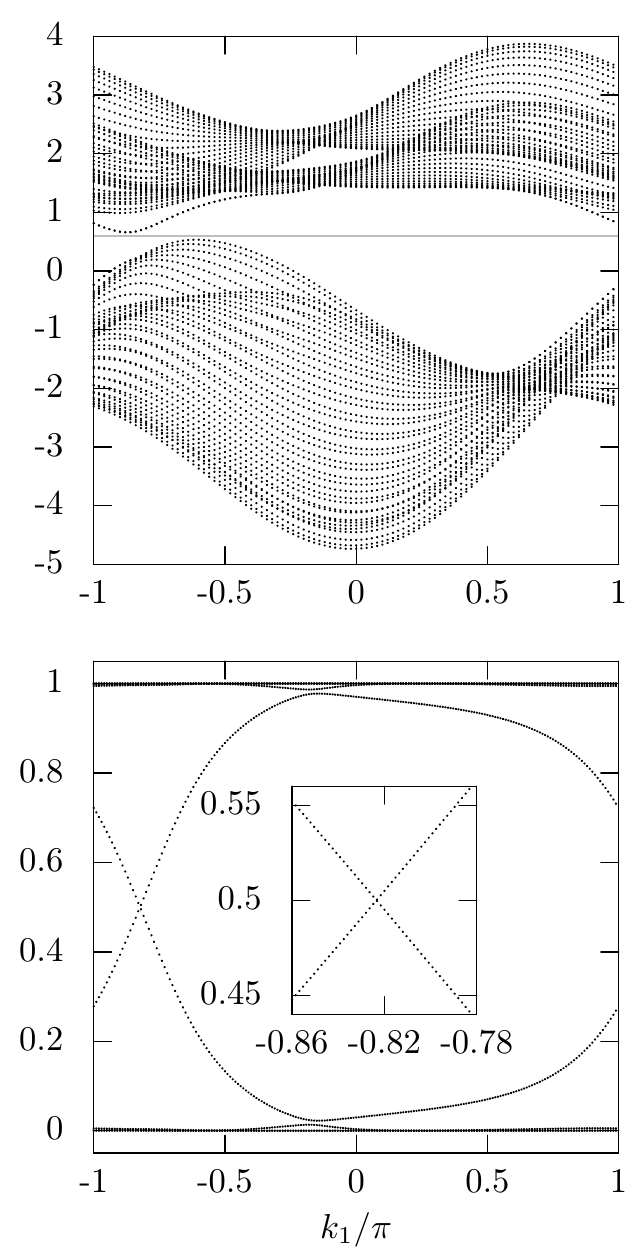}
  }
  \caption{Open boundary energy spectrum (upper panel) vs.
    entanglement occupancy spectrum (lower panel) of the bilayer
    model. Gray line indicates Fermi level. Parameters used are $[t_1,
    t_2, t_3] = [0.3,0.4,0.5]$, $m=1$, and $\tp = 0.5$. The Haldane
    phases $\phi$ and $\chi$ for each case are given in the subfigure
    title. (a) Both layers have Chern number $C=1$, giving a total
    $\sum C=2$. The entanglement occupancy spectrum has two flowing
    branches, each taking about one $k_1$ cycle ($k_1 \rightarrow k_1
    + 2\pi$) to flow from $f\simeq 1$ to $f \simeq 0$. (b) $AB$ layer
    has $C=1$ and $CD$ layer has $C=0$, hense $\sum C = 1$. Note the
    avoided crossing near $f \simeq 0.1$: If the two Haldane layers
    are not coupled, the crossing will not be avoided and one ends up
    having one flowing branch (which takes one $k_1$ cycle to flow
    from $f\simeq 1$ to $f\simeq 0$, see (a)) and another trivial
    branch. The avoided crossing effectively merges the two branches
    into a single branch which now takes about two $k_1$ cycles to
    flow from $f\simeq 1$ to $f\simeq 0$. (c) and (d): $AB$ and $CD$
    individually have opposite Chern indices. (c) Generic case where
    $\phi$ and $\chi$ are not related. (d) $\istar$-symmetric case
    with $\phi = -\chi$. Insets: details around half occupancy
    $f=\half$. In (d), the crossing at $f=\half$ is not lifted by the
    $\tp$ hopping. Crossings near $f=1$ and $0$ are avoided in both
    (c) and (d). Note that in all cases the number of gapless edge
    modes at a given boundary is given by $\sum C$, so is the number
    of entanglement spectral flows. (d) has no flow due to avoided
    crossing near $f = 0$ and $1$.}
  \label{blh-ergent}
\end{figure*}

A bilayer extension can be constructed by Bernal stacking two
monolayers with a vertical interlayer hopping $\tp$, as shown in
Fig.~\ref{blh}. $\tp$ couples any $A$ site with the $D$ site above it.
The Haldane phases of the two layers, $\phi$ for $AB$ and $\chi$ for
$CD$, are taken to be independent. Semenoff masses for the four
sites are $m_A = -m_B = -m_C = m_D = m$. Both layers have the same
second neighbor hoppings $t_i, i=1, 2, 3$. The bulk Hamiltonian is
thus
\begin{gather}
  \label{h-blh}
  H(\vec k) =
  \begin{pmatrix}
    H_{AB}(\vec k) & \tilde T\\
    \tilde T^{\rm T} & H_{CD}(\vec k)
  \end{pmatrix}\quad , \quad
  \tilde T =
  \begin{pmatrix}
    0 & \tp\\ 0 & 0
  \end{pmatrix}
\end{gather}
where $H_{AB}$ and $H_{CD}$ are given by eqns.~\ref{hbulk} and
\ref{omega-b} with different parameters $H_{AB} = H(\phi, m)$ and
$H_{CD} = H(\chi, -m)$. Superscript $t$ denotes matrix transposition.
The two layers individually can have different Chern indices. Thus as
long as the central gap does not close when $\tp$ is turned on, the
total Chern number should be given by the sum of those of the
individual layers. This is numerically verified, see
Fig.~\ref{blh-ergent}, where we plot open boundary energy spectrum and
entanglement occupancy spectrum.

An interesting case is when $\phi = -\chi$ (Fig.~\ref{blh-c0i}). The
bulk Hamiltonian is mapped to its complex conjugate under inversion,
\begin{gather}
  \label{istar}
  \mathcal{I} \, H(\vec k) \> \mathcal{I} = H^{*}(\vec k)\quad , \quad
  \mathcal{I} \equiv
  \begin{pmatrix}
    & & & 1\\
    & & 1\\
    & 1\\
    1
  \end{pmatrix}\ .
\end{gather}
We shall refer to this as the $\istar$ symmetry. If both layers are
individually topological, their Chern numbers are opposite
and there is no gapless edge mode. The entanglement occupancy
spectrum also exhibits no spectral flow, but unlike the generic $\phi
\neq -\chi$ case, there are protected half occupancy modes occuring at
the $k_1$ very close to the edge crossing $k_{\rm c}$ point of the
corresponding monolayer (clearly, they must be identical when $\tp =
0$). This is reminiscent of the $\mathbb{Z}_2$ inversion-symmetric
topological insulators\cite{hughes-prodan-bernevig11-inversion}, which
also shows protected $f=\half$ entanglement modes \emph{without} gapless
edge modes in the energy spectrum. An essential difference is that
there, they only occur at inversion-symmetric $k_1 = 0$ or $\pi$
points because inversion relates $H(\vec k)$ to $H(-\vec k)$.

Given the connection between entanglement spectrum and Wannier
centers, the question naturally arises as to how the latter would
behave in the presence of $\istar$ symmetry. We shall address this
question in this section.

\subsection{Non-Abelian Wannier centers and Wilson loop phases}
The Haldane bilayer has $\nu = 2$ bands filled at half filling. For adiabatic
evolution of $\nu$ bands, the concept of Berry connection
(``geometrical vector potential'') generalizes to a $\nu \times \nu$
Berry connection matrix (``non-Abelian gauge field'')\cite{WZ84},
\begin{gather}
  \hat A_{ab} = \langle \psi_k^a | \, i\,\partial_k \, | \psi_k^b\rangle
\end{gather}
where $|\psi_k^a\rangle$ is the internal-space eigenstate at $k$ of
the $a^{\rm th}$ band, \emph{etc.} The Wilson loop, also a $\nu \times
\nu$ matrix, is defined as
\begin{gather}
  \ww = \wwp \exp\left\{i\!\int \limits_0^{2\pi} \!\! dk\, \hat A(k) \right\}
\end{gather}
where $\wwp$ denotes path ordering. The phases of its eigenvalues play
the role of Berry phase.

The Wannier states and Wannier centers can still be defined as the
eigenstates and eigenvalues of the band-projected position operator
eqn.~\ref{grg}, where $\g$ now consists of $\nu$ bands, $\g = \sum_{a
  = 1}^{\nu} \sum_{k} |\Psi_{k}^a\rangle\langle \Psi_{k}^a|$, and
$|\Psi_{k}^a\rangle~=~|k\rangle~\otimes~|\psi_{k}^a\rangle$ is
eigenstate of the $a^{\rm th}$ band below Fermi level. One then finds that
as the number of lattice sites $N \rightarrow \infty$, the Wannier
centers are given by \cite{Qi11-wannier}
\begin{gather}
  \label{wloop-phase}
  x_{w,I} = {\gamma_w\over 2\pi} + I
\end{gather}
where $I$ is an integer labeling the corresponding unit cell, and
$\gamma_w$ is given by the phase of the $w^{\rm th}$ eigenvalue of $\ww$,
\begin{gather}
  \ww \, \mathbf{g}_w = e^{i\gamma_w} \, \mathbf{g}_w\quad , \quad w = 1, 2,
  \ldots, \nu \ .
\end{gather}
The $\nu$ bands recombine to yield $\nu$ Wannier functions associated
with each unit cell. When $\nu = 1$, this reduces to
eqn.~\ref{wannier-1b}.

In the finite $N$ case, it is more convenient to replace the position
operator $R = \hat Y \otimes \mathbb{I}$ with the $k$-space
translation operator $\exp(i \hat Y \delta k)\otimes \mathbb{I}$ where
$\delta k = 2\pi / N$ is the $k$-space step \cite{Yu11-z2-wannier}.
The object of concern is instead (cf.~derivation of eqn.~\ref{monodromy-app})
\begin{gather}
  \label{monodromy}
  \wwb = P_{k\nd_N} P_{k\nd_{N-1}} P_{k\nd_{N-2}} \cdots
  P_{k\nd_2} P_{k\nd_1} P_{k\nd_N}\,.
\end{gather}
where $k_n = n\, \delta k$ with $n = 1, 2, \ldots, N$, and $ P_{k_n} =
\sum_{a = 1}^{\nu} |\psi_{k_n}^a\rangle \langle \psi_{k_n}^a|$ is the
internal-space projector onto the occupied bands at $k$. Note that
this object is basis-independent---in particular, it is oblivious to
the phase choice of the $|\psi_k^a\rangle$ states, which is
advantageous for numerics. Note also that the dimension of $P_k$ and
hence of $\wwb$ is the number of \emph{total} bands $q$ whereas that
of $\ww$ is $\nu$, the number of \emph{occupied} bands. However, $\ww$
is essentially the nonzero block of $\wwb$ (cf.~eqn.~\ref{wsub}),
hence they have the same (complex) nonzero eigenvalues $\{\rho_w\,
e^{i\gamma_w}\}$ where $w = 1, 2, \ldots, \nu$. The amplitudes
$\rho_w$ will deviate from unity due to discretization. The phases
$\gamma_w$ will be interpreted as the Wannier center offsets as in
eqn.~\ref{wloop-phase}. See Appendix \ref{wcwl} for details.

$\wwb$ in the form of eqn.~\ref{monodromy} -- as a product of occupied
band projectors over the period of $k$ -- is also known as a monodromy
and has been used in analyzing \emph{e.g.} inversion-symmetric
$\mathbb{Z}_2$ topological insulators
\cite{hughes-prodan-bernevig11-inversion}. One question is whether or
not it matters to break up the Wilson loop at some $k$ point other
than $k=0$. The answer is no. This is because the eigenvalues of a
product of (possibly non-commuting) projectors are invariant under
cyclic permution of these projectors. We prove it in
Appendix~\ref{pprod}.

\begin{figure*}[t!]
  \subfigure[\;$\phi = 0.3\pi$, $\chi = -0.4\pi$]{
    \label{blh-wan-c0}\includegraphics[width=0.45\textwidth]{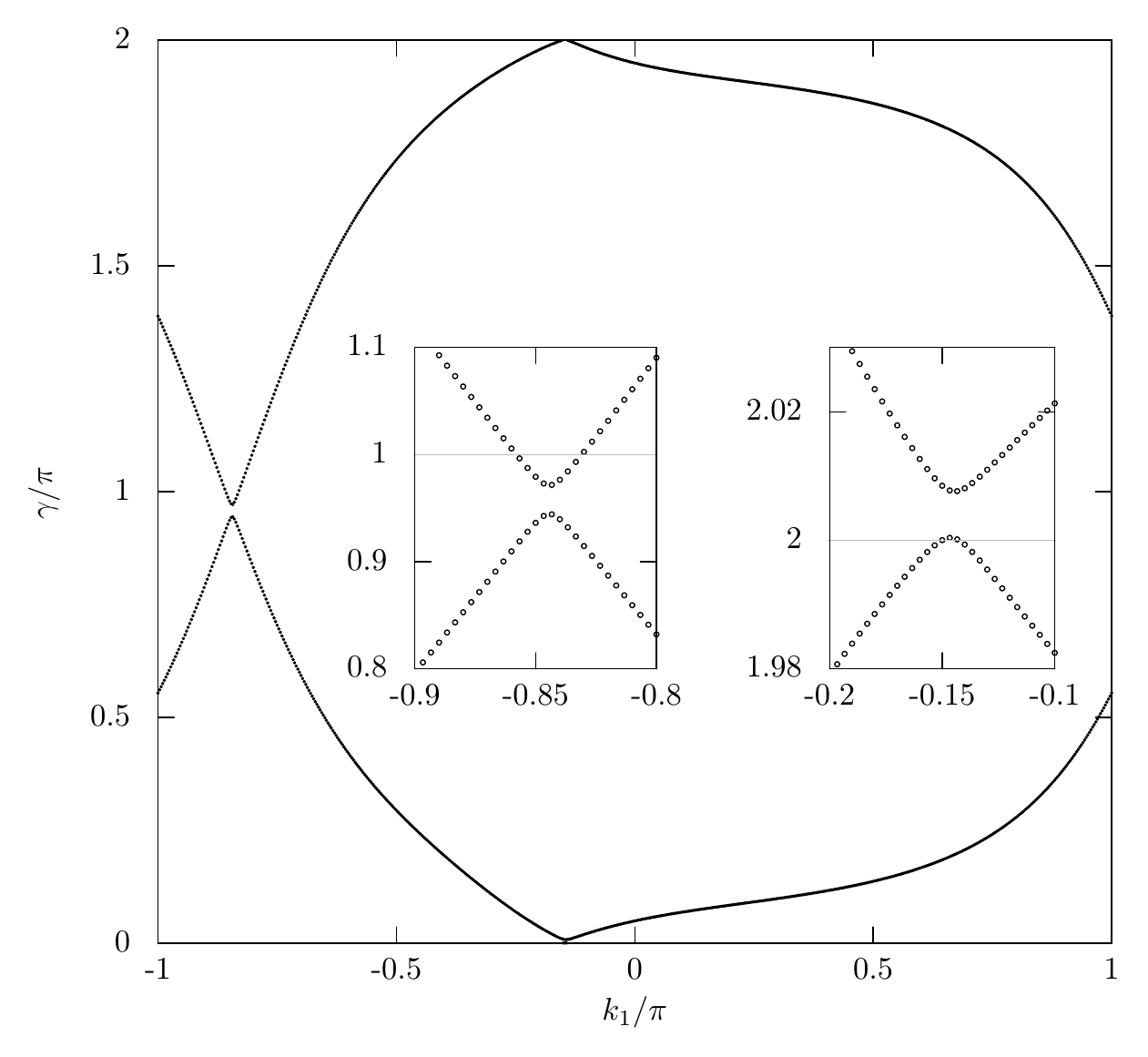}
  }
  \subfigure[\;$\phi = 0.3\pi$, $\chi = -0.3\pi$]{
    \label{blh-wan-c0i}\includegraphics[width=0.45\textwidth]{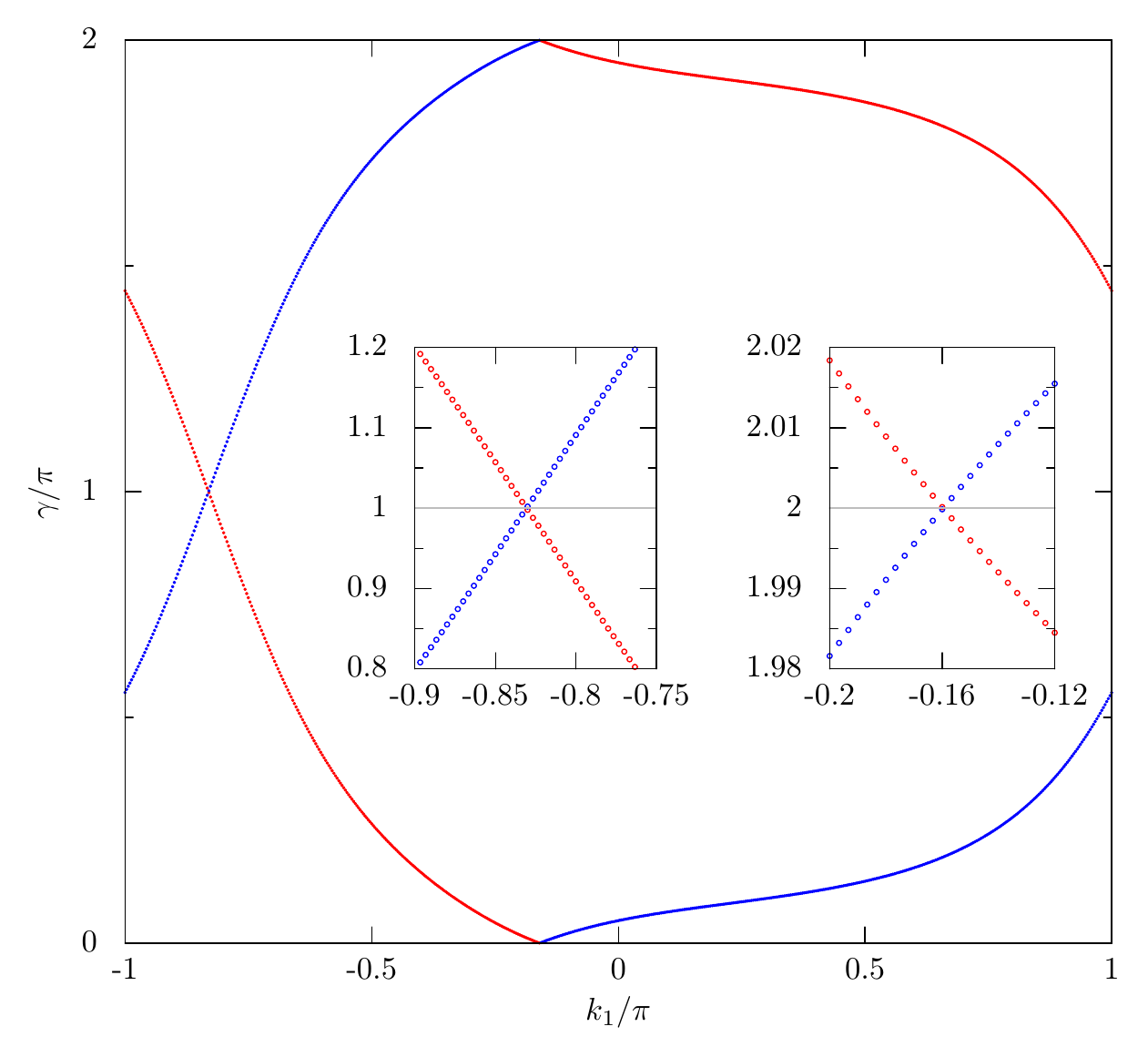}
  }
  \caption{(Color online) Non-abelian Wilson loop phases. Haldane
    phases are given in subfigure titles. Other parameters used are
    the same as in Fig.~\ref{blh-ergent}. Insets: details around
    $\gamma = \pi$ and $2\pi$. (a) has avoided crossings near both
    $\gamma = \pi$ and $2\pi$, but (b) has no avoided crossings. The
    red dots in (b) can be computed by eqn.~\ref{blhi-gamma}. The blue
    dots are given by their negatives (shifted by $2\pi$).}
  \label{blh-wan-compare}
\end{figure*}

\subsection{Wilson loop phase of the $\istar$-symmetric bilayer Haldane model}

\subsubsection{$\istar$-symmetric bilayer eigenstates}
Upon swapping the $C$ and $D$ rows and columns, the Hamiltonian
eqn.~\ref{h-blh} becomes (suppressing $k_1$ dependence)
\begin{gather}
  \label{h-blhi}
  H = \omega \, \mathbb{I} + 
  \begin{pmatrix}
    M & T\\
    T^{*} & M^{*}
  \end{pmatrix}
\end{gather}
\begin{gather}
  \label{mt}
  M = \vec B \cdot \vec \sigma\quad,\quad
  T = T^{*} = 
  \begin{pmatrix}
    \tp & 0\\ 0 & 0
  \end{pmatrix}\ ,
\end{gather}
where $\omega$ and $\vec B$ are given by eqn.~\ref{omega-b}. The form
of eqn.~\ref{h-blhi} means its eigenstates are of the form
\begin{gather}
  |\psi\rangle =
  \begin{pmatrix}
    |u\rangle \\ |u^*\rangle
  \end{pmatrix}
\end{gather}
where $|u\rangle = \bigl(
\begin{smallmatrix}
  u_1\\u_2
\end{smallmatrix}
\bigr)$ is a two-component column vector. The eigenvalue equation is
thus
\begin{gather}
  M |u\rangle + T |u^*\rangle = \varepsilon |u\rangle\ .
\end{gather}
Since interlayer hopping is restricted to $A$ and $D$ sites, $T
|u^*\rangle$ only depend on $u_1$, thus one may always adopt a phase
choice of $|u\rangle$ such that $u_1$ is either real, in which case $T
|u^*\rangle = T |u\rangle$, or imaginary, in which case $T
|u^*\rangle = - T |u\rangle$. The eigenvalues of $H$ consequently
comprise those of $2\times 2$ $H^{\pm}$ defined as
\begin{gather}
  H^{\pm} \equiv \omega \, \mathbb{I} + M \pm T = \omega_{\pm} \, \mathbb{I} + \vec B_{\pm} \cdot \vec \sigma\ ,\\
  \omega_{\pm} = \omega \pm \half\tp\ ,\\
  \vec B_{\pm} = \bigl(B_x, B_y, B_z \pm \half\tp\bigr) =
  \big(|B_{\pm}|, \vartheta^{\pm}, \varphi\big)
\end{gather}
where $\big(|B_{\pm}|, \vartheta^{\pm}, \varphi\big)$ are the spherical polar
coordinates of $\vec B_{\pm}$. The role of $\tp$ is two-fold: by
modifying $\omega$, it splits the two monolayer copies; by modifying
$B_z$, it also changes the level splitting of each monolayer. Thus as
$\tp$ is increased from $0$ adiabatically, it is possible to rearrange
the order of the monolayer bands. Here we shall focus on the case
where the central gap is never closed during $0 \rightarrow \tp$, so
that the occupied bands $|\psi^{\pm}(\vec k)\rangle$ at half filling
are generated by the isospin-down states of $H^{\pm}$, \emph{viz.},
\newcommand{\thp}{\vartheta^{+}} \newcommand{\thm}{\vartheta^{-}}
\begin{gather}
  \notag
  |\psi^{+}\rangle = \frac{i}{\sqrt{2}} D_z(\varphi)
  \begin{pmatrix}
    \phantom{+}|\thp\rangle\\
    -|\thp\rangle
  \end{pmatrix}\ ,\\
  \label{psi-pm}
  |\psi^{-}\rangle = \frac{1}{\sqrt{2}} D_z(\varphi)
  \begin{pmatrix}
    |\thm\rangle\\
    |\thm\rangle
  \end{pmatrix}\ ,
\end{gather}
where $D_z(\varphi) = \textsf{diag}(1,e^{i\varphi}, 1, e^{-i\varphi})$
and $|\vartheta^{\pm}\rangle = 
\begin{pmatrix}
  -\sin \frac{\vartheta^{\pm}}{2}\\
  \cos \frac{\vartheta^{\pm}}{2}
\end{pmatrix}$ are purely real. Note that $|\psi^{\pm}\rangle$ have the prescribed
form $\begin{pmatrix}
  |u\rangle^{\phantom *} \\ |u^*\rangle\end{pmatrix}$
  with $u_1$ imaginary for $|\psi^{+}\rangle$ and real for $|\psi^{-}\rangle$.

\subsubsection{Wilson loop phases}

Fig.~\ref{blh-wan-compare} plots the Wilson loop phases of the Haldane bilayer
where the two layers have opposite Chern indices. In the generic $\phi
\neq -\chi$ case (Fig.~\ref{blh-wan-c0}), both $\gamma_w$ levels
exhibit no flow over the period of $k_1$ due to avoided crossings,
which is similar to the corresponding entanglement spectrum
(Fig.~\ref{blh-c0}). In the $\istar$-symmetric case where $\phi =
-\chi$, none of the crossings at $\gamma = 0$ and $\pi$ are avoided,
thus both $\gamma_w$ levels flow in opposite directions. This is
different from the entanglement spectrum (Fig.~\ref{blh-c0i}) where
the crossing at $f = \half$ is preserved, but the crossings near $f = 1$
and $0$ are avoided. Unlike the monolayer case, the $f=\half$ crossing
does not coincide in $k_1$ with the $\gamma = \pi$ crossing, although
they are very close. This is because one no longer has the analog of a
coplanar $k_1$ point. We will come back to this point later.

Since the set $\{\gamma_w\}$ is periodic as $k_1 \rightarrow k_1 +
2\pi$, their winding numbers $\bigl[\gamma_w(2\pi) -
\gamma_w(0)\bigr]/(2\pi)$ must be integers. The question is how to
extract them. From eqn.~\ref{psi-pm}, one finds that the Berry
connection matrix is purely off-diagonal,
\begin{align}
  \langle \psi^{+} | \partial_{k\nd_2} | \psi^{+}\rangle &=
  \langle \psi^{-} | \partial_{k\nd_2} | \psi^{-}\rangle = 0\,,\\
  \langle \psi^{-} | \partial_{k\nd_2} | \psi^{+}\rangle &=
  - {\partial \varphi\over \partial k_2} \, \cos(\half\thp) \cos(\half\thm)\,,
\end{align}
hence
\begin{gather}
  \ww(k_1) = e^{i\tilde{\gamma}(k_1) \, \sigma_y}\,\\
  \tilde{\gamma}(k_1) = -\!\!\!\int\limits_{\varphi(0)}^{\varphi(2\pi)}\!\!\!
  d\varphi\, \cos \frac{\thp}{2} \cos \frac{\thm}{2}\ .
  \label{blhi-gamma}
\end{gather}
Note that when $\tp = 0$, $\thp = \thm$ and eqn.~\ref{blhi-gamma}
reduces to eqn.~\ref{mlh-gamma}. The Wilson loop phases are simply
$\pm \tilde{\gamma}$. In Fig.~\ref{blh-wan-c0i}, red curve corresponds
to $+\tilde{\gamma}$ and blue to $-\tilde{\gamma}$.

While there are no inversion or time-reversal symmetries protecting
the winding of this off-diagonal $\tilde{\gamma}$ (by fixing
$\tilde\gamma = 0$ and $\pi$ at $\mathcal{T}$ or
$\mathcal{I}$ symmetric $k_1$ points), it is nontheless robust with
respect to $\tp$ as long as the central gap remains open, and is thus
given by the corresponding monolayer Chern number ($\tp = 0$ case). To
see this, we use the eigenstates of $\ww$ (cf.~eqns.~\ref{lwi} and
\ref{fwi}) to construct a set of states,
\begin{align}
  |\eta_{\pm}\rangle &= g_{\pm}^{+} \,|\psi^{+}\rangle + g_{\pm}^{-}\,
  |\psi^{-}\rangle = \frac{1}{\sqrt{2}}\big\{|\psi^{+}\rangle \pm i
  |\psi^{-}\rangle \big\}\\
  &= \frac{i}{2} D_z(\varphi)
  \begin{pmatrix}
    |\thp\rangle + |\thm\rangle\\
    |\thm\rangle - |\thp\rangle
  \end{pmatrix}
\end{align}
where $\mathbf{g}_{\pm} = (g_{\pm}^{+}, g_{\pm}^{-})^{\rm T}$ are the
eigenstates of $\ww$, in this case, the eigenstates of $\sigma_y$. The
Berry connection matrix $\hat A$ is diagonal in the
$|\eta_{\pm}\rangle$ basis, \emph{i.e.}, $\tilde\gamma$ is the Berry
\emph{phase} (over $k_2$) of the $|\eta_{+}\rangle$ ``band''. It is
easy to verify that the two layers, $AB$ and $CD$, each contributes
exactly one half to $\tilde \gamma$. One may thus construct a
ficticious two-band model whose occupied band is given by the
projection of say $|\eta_{+}\rangle$ onto the $AB$ layer,
\begin{gather}
  \label{mstate}
  |-\rangle = \frac{1}{\sqrt{2}}
  \begin{pmatrix}
    1\\ & e^{i\varphi}
  \end{pmatrix}
  \bigl(|\thp\rangle + |\thm\rangle \bigr)\ ,
\end{gather}
and the unoccupied band as
\begin{gather}
  |+\rangle = \frac{1}{\sqrt{2}}
  \begin{pmatrix}
    1\\ & e^{i\varphi}
  \end{pmatrix}
  \bigl(|\bar\vartheta^{+}\rangle - |\bar\vartheta^{-}\rangle\bigr)\ .
\end{gather}
Here, $|\bar\vartheta\rangle = \begin{pmatrix}\cos(\vartheta/2)\\ \sin(\vartheta/2)
\end{pmatrix}$ are the iso-spin up counterpart of $|\vartheta\rangle$. By
construction, $\langle + | - \rangle = 0$. The ficticious Hamiltonian
is
\begin{gather}
  \tilde H = \sum_{\vec k, s = \pm}\varepsilon_s(\vec k) \, |\vec
  k\rangle \langle \vec k| \otimes |s(\vec k)\rangle\langle s(\vec k)|
\end{gather}
where $\varepsilon_{+}$ can be chosen as (say) the lower energy of the
two unoccupied bilayer bands, and $\varepsilon_{-}$ as the higher energy
of the two occupied bilayer bands. Then $\tilde \gamma(k_1)$ is the Berry
phase of $\tilde H$, and its winding number is the same as the
monolayer Haldane model as long as $\tp$ does not collapse the bilayer
central gap. The reason why the point of $\tilde \gamma = \pi$ is very
close to but no longer coincide with where the $f = \half$ entanglement
mode occurs, which was the case for monolayers, is that if one maps
the state of eqn.~\ref{mstate} back to a vector $\vec B$, with its
azimuth as $\varphi$ and its polar angle being some sort of average of
$\thp$ and $\thm$, then around the monolayer $k_{\rm c}$, the path of $\vec
B$ comes close to coplanarity (yet never exactly so). Since the
``temperature'' $\beta$ is different for the entanglement and Wannier
spectra (cf.~eqn.~\ref{rbeta-eqn}), they need different distortions in
the $\vec B$ path to balance $\beta$ in order to reach their special
values.

\begin{figure*}[t!]
  \subfigure[\;$\phi = 0.3\pi, \chi = -0.4\pi, N_2 = 500$]{
    \label{blhex-c0}\includegraphics[width=0.45\textwidth]{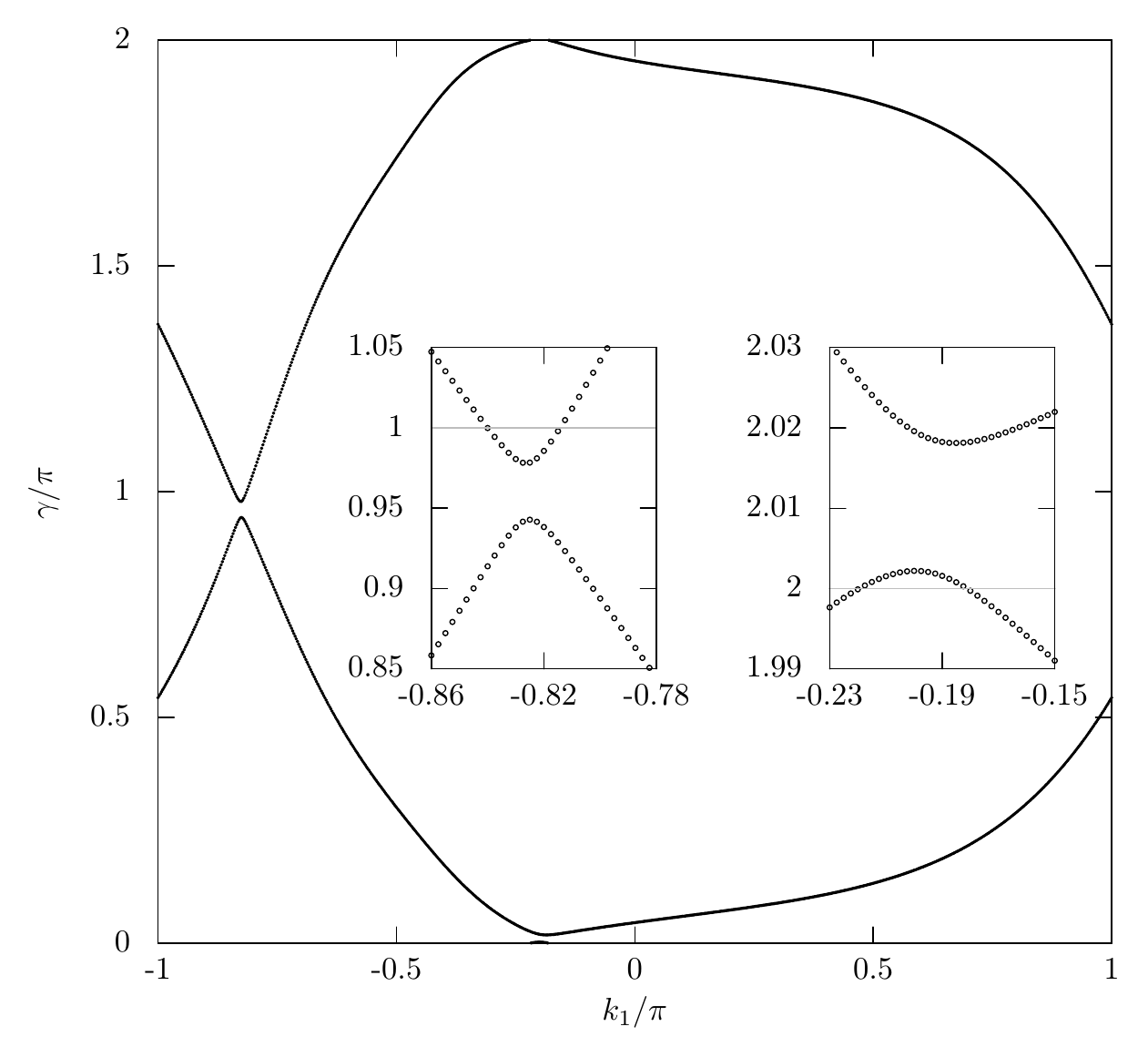}}
  \subfigure[\;$\phi = 0.3\pi, \chi = -0.3\pi, N_2 = 500$]{
    \label{blhex-c0i}\includegraphics[width=0.45\textwidth]{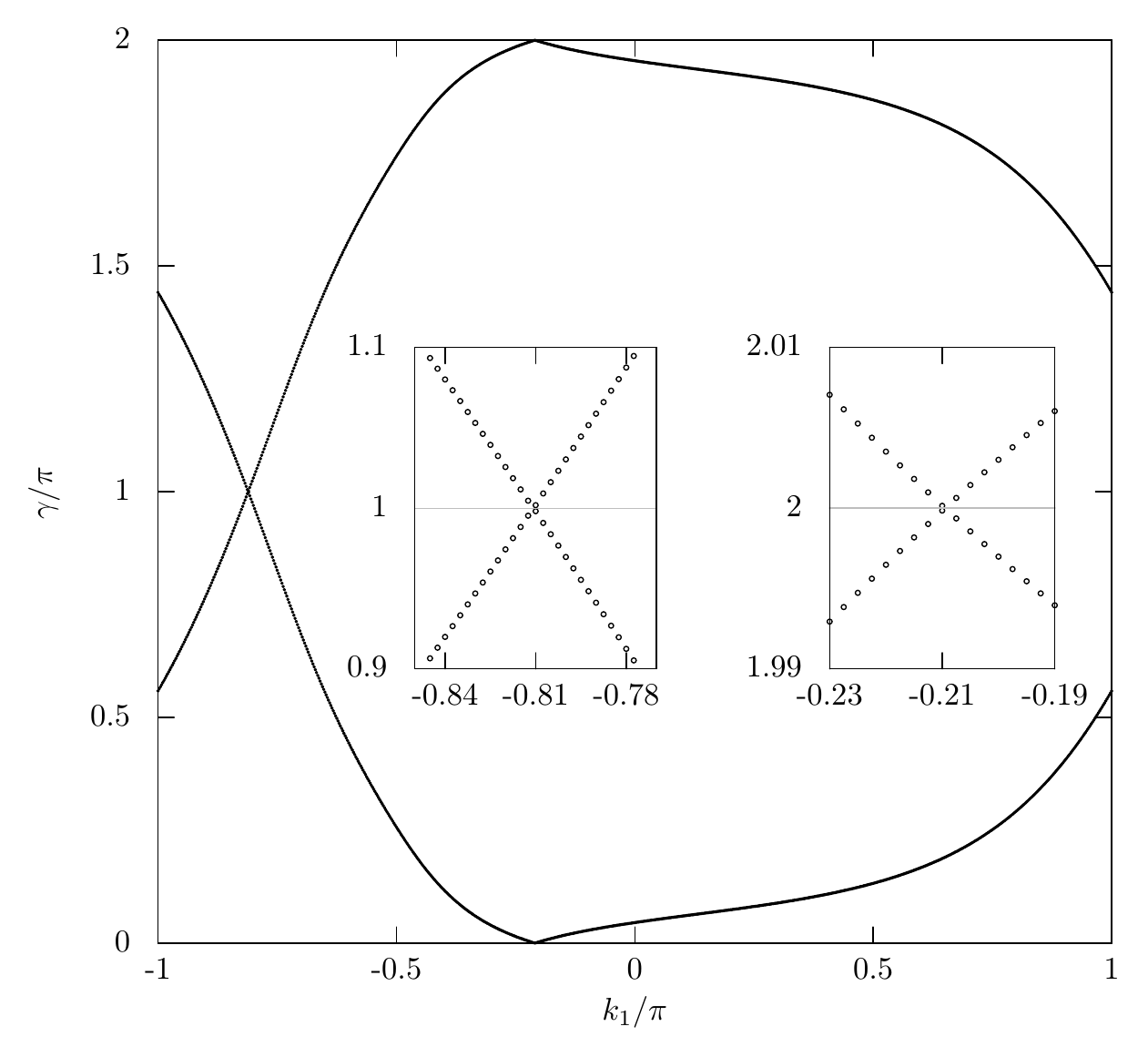}}\\
  \subfigure[\;$\phi = 0.3\pi, \chi = -0.4\pi, N_2 = 10$]{
    \label{blhex-c0-undersampling}\includegraphics[width=0.45\textwidth]{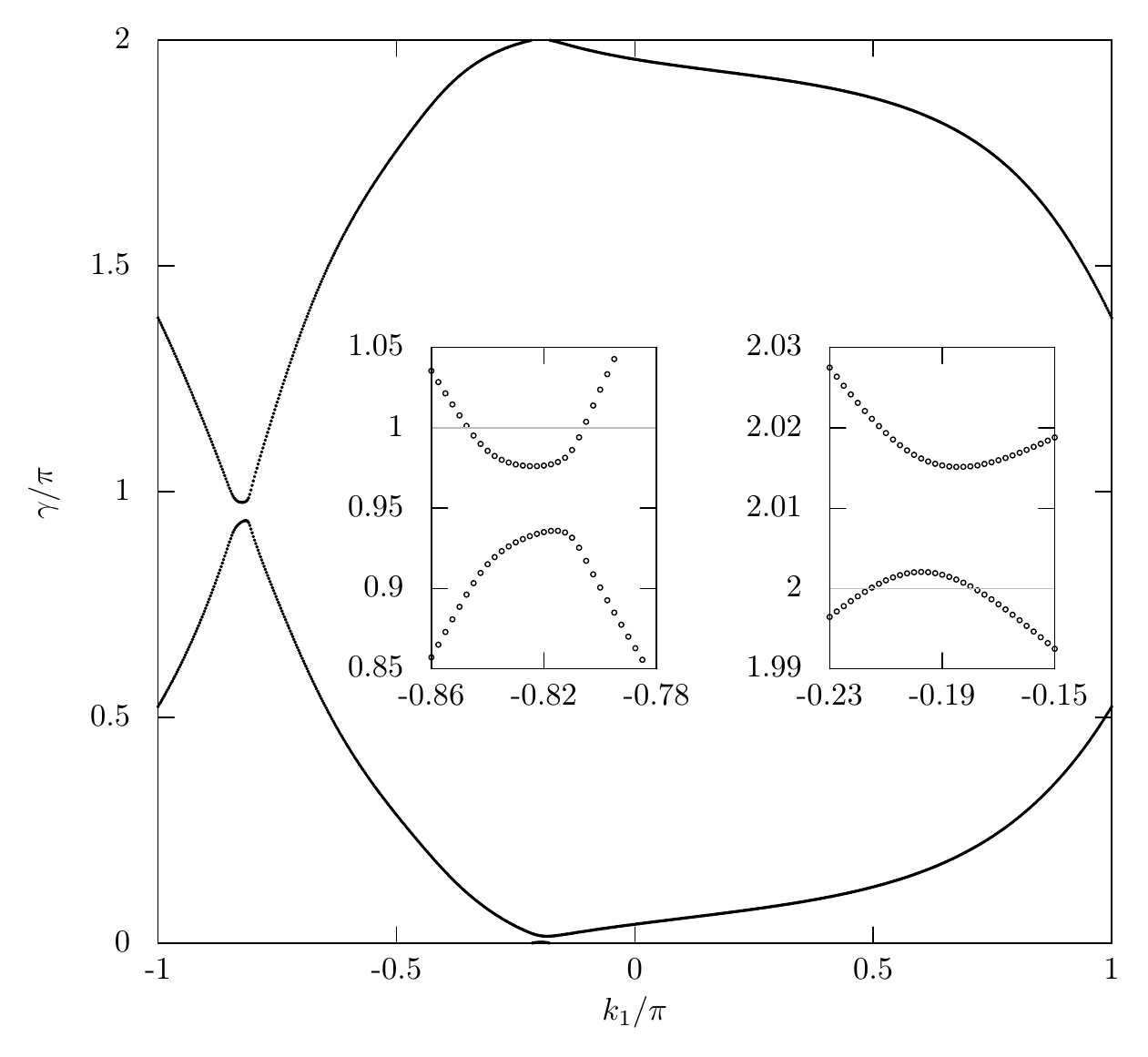}}
  \subfigure[\;$\phi = 0.3\pi, \chi = -0.3\pi, N_2 = 10$]{
    \label{blhex-c0i-undersampling}\includegraphics[width=0.45\textwidth]{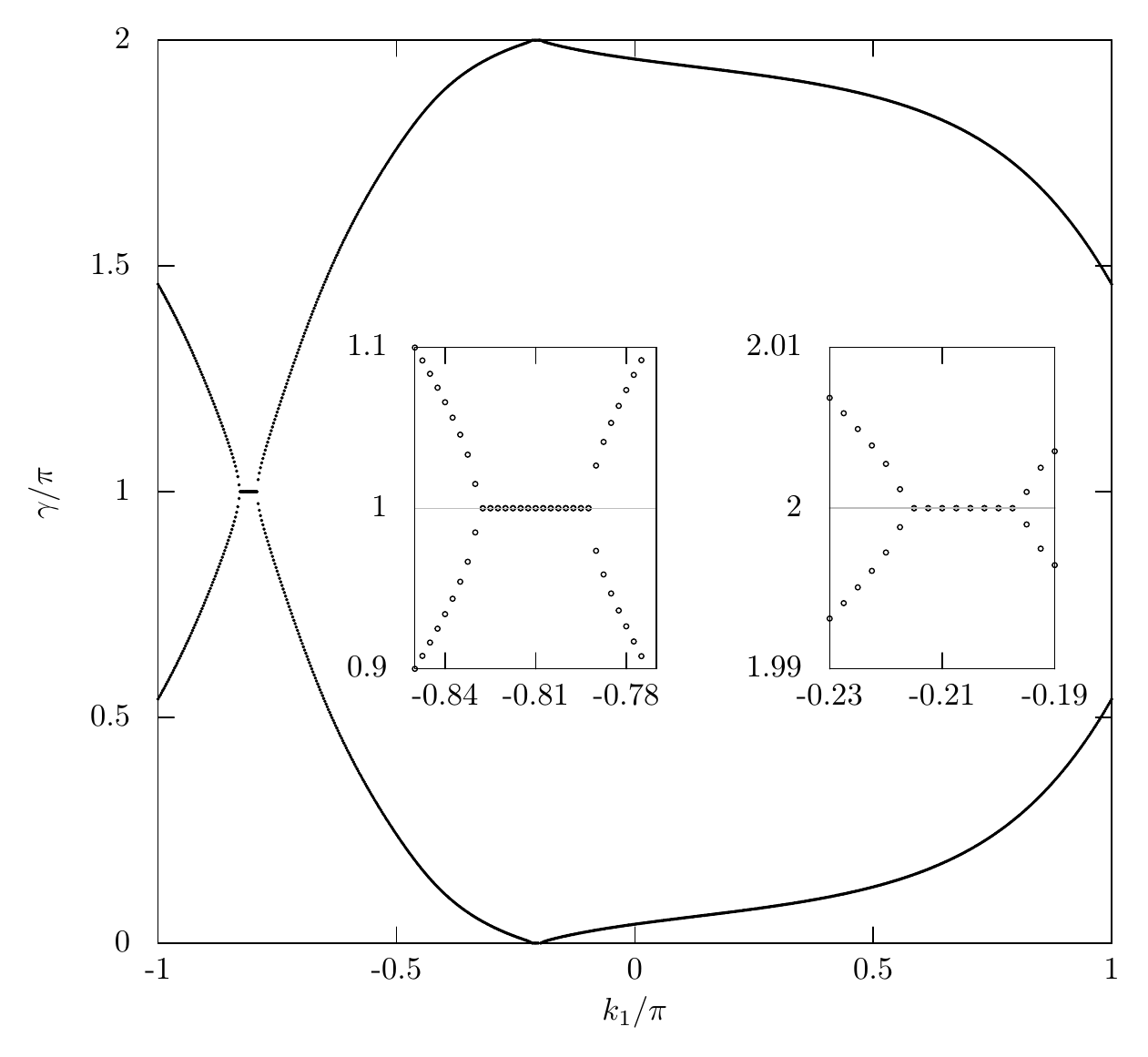}}
  \caption{Wilson loop phases of bilayer Haldane model with generic
    $\istar$-preserving interlayer hopping, and the effect of $k_2$
    undersampling. Parameters used are $\tp = 0.5$, $\lambda = 0.4$,
    other parameters are the same as in Fig.~\ref{blh-ergent}. Insets:
    details around $\gamma = \pi$ and $2\pi$. (a): $\phi \neq -\chi$,
    \emph{i.e.}, $\istar$-non-symmetric. (b): $\phi = -\chi$,
    \emph{i.e.}, $\istar$-symmetric. (c): same as (a) but only use
    $N_2 = 10$ projectors in eqn.~\ref{monodromy}. (d): same as (b)
    but with $N_2 = 10$. Notice that qualitative features do not
    depend on the number of $k_2$ discretization. In particular, $k_2$
    undersampling does \emph{not} lift the crossing in the
    $\istar$-symmetric case.}
\end{figure*}

\subsection{Generic $\istar$-preserving interlayer hopping}
The reason why $\tilde\gamma$ can be extracted from the bilayer Haldane model is
because the Berry connections $\hat A(\vec k)$ are proportional to
$\sigma_y$, and hence are mutually commuting at different $\vec k$.
This in turn is because the interlayer hopping is only between $A$ and
$D$ sites, so that $H$ decomposes into $H^{\pm}$ sectors. Thus while
there is a matrix structure, the situation is nontheless Abelian.
Still, the $\istar$-symmetric bilayer provides an interesting
example where the topological information is not obvious from the
open-boundary energy spectrum. The role of the $\istar$ symmetry is in
prescribing the eigenstates to a particular form $\bigl(
\begin{smallmatrix}
  |u\rangle^{\phantom *} \\ |u^*\rangle
\end{smallmatrix}
\bigr)$ (after swapping $C$ and $D$ labeling). One question to ask is
if the crossings as seen in the Wilson loop phases of the
$\istar$-symmetric bilayer is a consequence solely of the $\istar$
symmetry, or if it also depends on the interlayer hopping being
restricted to $\tp$ only. The most general interlayer hopping that
preserves $\istar$ is
\begin{gather}
  \label{t-istar}
  T =
  \begin{pmatrix}
    \lambda_1 & \lambda_2\\
    \lambda_2 & \lambda_3
  \end{pmatrix}\quad , \quad \lambda_i \in \mathbb{C}\ .
\end{gather}
instead of the one used in eqn.~\ref{mt}. In the bilayer stacking, the
$B$ sites are surrounded by three ``nearest-neighbor'' $C$ sites on
the other layer, $C$ by $A$, and $D$ by $B$, thus one example of such
$T$ matrix is by treating the strength of all these interlayer
hoppings as $\lambda$ in addition to the vertical $\tp$ between $A$
and $D$,
\begin{gather}
  \notag
  \lambda_1 = \tp\ ,\\
  \notag
  \lambda_2 = \lambda(1 + e^{ik_1} + e^{ik_2})\ ,\\
  \lambda_3 =  \lambda(e^{i k_1} + e^{i k_2} + e^{i (k_1 + k_2)})\ .
\end{gather}
Although not important for our purpose here, these parameters can be
connected to the SWMC model of graphene\cite{nilsson08-mgraphene}.

$\istar$ symmetry of the Hamiltonian is inherited by the
internal-space projector of occupied bands, and hence their product
$\wwb$,
\begin{gather}
  \mathcal{I} P(\vec k) \,\mathcal{I} = P^{*}(\vec k)\quad , \quad
  \mathcal{I} \,\wwb (k_1) \,\mathcal{I} = \wwb^{*}(k_1)\ .
\end{gather}
So if $|g_w\rangle$ is an eigenstate of $\wwb$ with eigenvalue
$\rho_w \, e^{i\gamma_w}$, then $\mathcal{I}\,|g_w\rangle^{*}$ is also an
eigenstate with eigenvalue $\rho_w \, e^{-i\gamma_w}$. Thus the
eigenvalues of $\wwb$ come in complex-conjugate pairs. For the bilayer
with a general interlayer hopping eqn.~\ref{t-istar}, while there does
not seem to be a way of tracking one of the two $\wwb$ eigenstates
(with nonzero $\rho_w$) analytically, we do find numerically that the
$\{\gamma_w\}$ pair flow in opposite directions and cross at $\gamma =
0$ and $\pi$ just like the case with $\tp$-only interlayer hopping,
and that their winding numbers are given by the underlying monolayer
as long as the bilayer central gap does not collapse when turning on
$\tp$ and $\lambda$. Fig.~\ref{blhex-c0i} plots a $\istar$-symmetric
case with $T$ given by $\tp = 0.5$ and $\lambda = 0.4$. Compare with
Fig.~\ref{blhex-c0} where $\phi \neq -\chi$ and the flows are broken
near $\gamma = \pi$ and $2\pi$.

An interesting observation made possible by expressing the Wilson loop
as a product of projectors is how the topological and non-topological
cases behave with an undersampling of $\kpara$, that is, when the
number of projectors used in eqn.~\ref{monodromy} is reduced. For
example, if one were to reduce $N$ projectors to say $N/3$, it is
equivalent to replacing every two out of three projectors in the
monodromy by a unity operator, thereby allowing wavefunctions to leak
into \emph{unoccupied} states when ``propagating'' from $k_i$ to
$k_{i-3}$. Of course $\ww$ ceases to be unitary due to the leakage,
which is just the reason why $\rho_w < 1$ in the finite $N$ case. A
somewhat unexpected behavior is that the Wilson loop phases in the
$\istar$-symmetric case now become degenerate for an extended region
of $k_1$ values at $\gamma = \pi$ and $0$, as opposed to crossing at
discrete points in the $N\rightarrow \infty$ limit. The eigenvalues
there are simply real numbers, which are their own conjugates and thus
not ruled out by the $\istar$ symmetry. Note however that the two real
solutions are not required to be the same---indeed they are different
for finite $N$ and only approaches $\pm 1$ as $N \rightarrow \infty$
per unitarity of $\ww$. No such phase degeneracy is observed for the
generic $\istar$-asymmetric cases, where the phase flows still
exhibit avoided crossing. The topological signature is in this sense
more prominent away from the thermodynamic limit, unlike \emph{e.g.}
the crossing of the monolayer energy edge states which in general is a
thermodynamic limit result.

\subsection{Relation with a $\mathbb{Z}_2$ inversion-symmetric TI}
\begin{figure}[t!]
  \centering
  \includegraphics[width=0.45\textwidth]{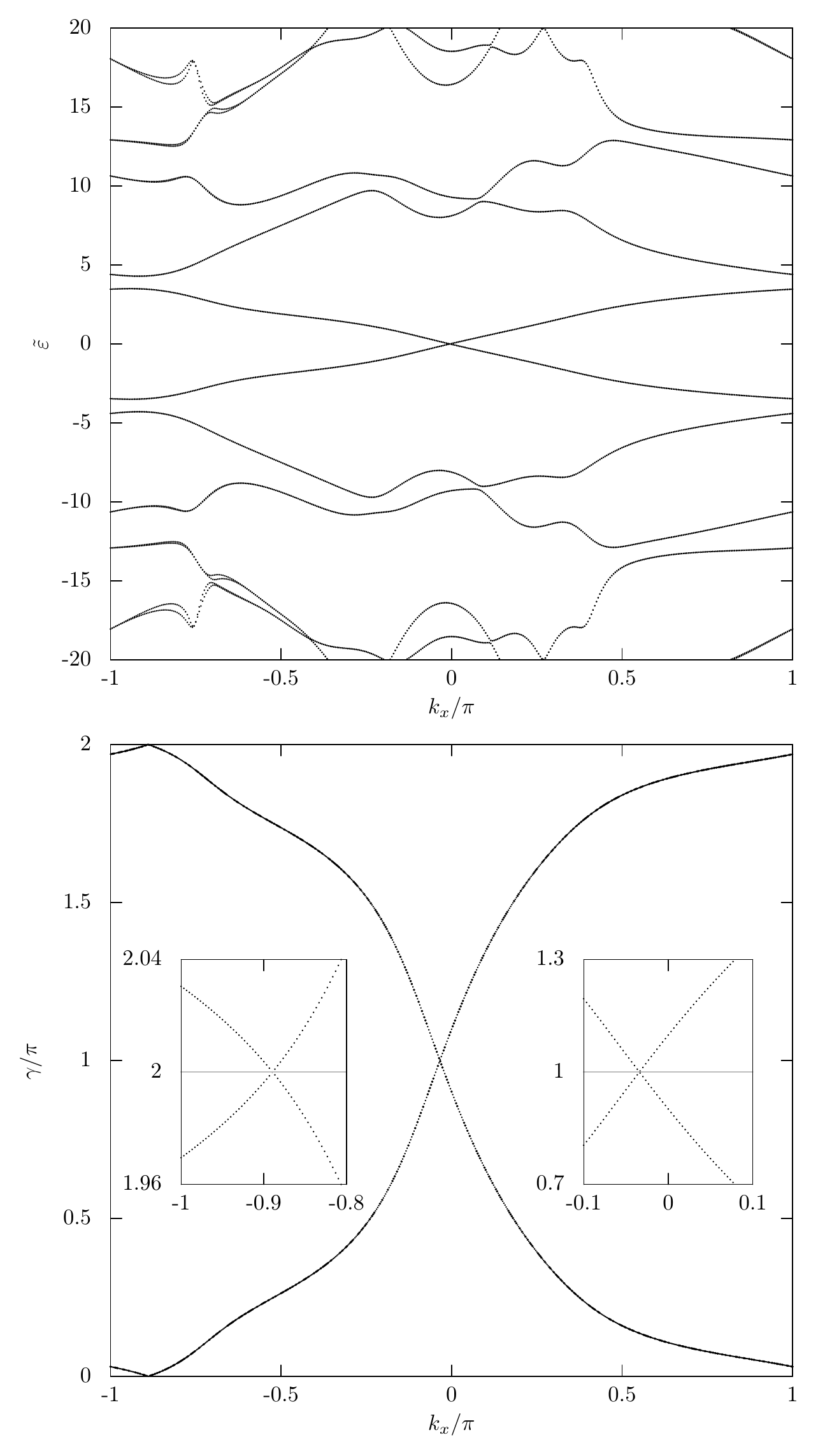}
  \caption{$\istar$-symmetric HPB model with broken inversion. Top
    panel: periodic boundary entanglement quasi-energy. Bottom panel:
    Wilson loop phases. Parameters: $m = 1.1$, $(\lambda_1, \lambda_2,
    \lambda_3) = (0.7, 0.5, 0)$. Since the $\istar$ symmetry is
    preserved, the model still retains its nontrivial topology: in the
    entanglement spectrum, although there is no spectral flow, the
    $\tilde \varepsilon = 0$ mode, which now is slightly shifted away
    from $k_x=0$, is still robust with no avoided crossing; for the
    Wilson loop phases, the two branches still wind in opposite
    directions and intersect at $\gamma = 0$ and $\pi$, but the
    crossings are no longer pinned at $k_x = 0$ or $\pi$. Insets:
    details near $\gamma = \pi$ and $2\pi$. }
  \label{hpbmod}
\end{figure}

Hughes \emph{et.~al.~}studied a $\mathbb{Z}_2$ inversion-symmetric TI
model (HPB model) \cite{hughes-prodan-bernevig11-inversion},
\begin{gather}
  \label{hpb-h}
  H(\vec k) = \sin k_x \Gamma_1 + \sin k_y \Gamma_2 + M(\vec k) \Gamma_0 + B_x \Gamma_B\ ,\\
  M(k) \equiv 2 - m - \cos k_x - \cos k_y\ ,\\
  \Gamma_1 = \sigma_z \otimes \tau_x\quad , \quad \Gamma_2 = \mathbb{I} \otimes \tau_y\ ,\\
  \Gamma_0 = \mathbb{I} \otimes \tau_z\quad , \quad \Gamma_B =
  \sigma_x \otimes \left(
    \begin{smallmatrix}
      1\\ & 0
    \end{smallmatrix}\right)\ .
\end{gather}
Here, both $\sigma_i$ and $\tau_i$ are the Pauli matrices, $\sigma_i$
act on the spin indices and $\tau_i$ on the orbital indices. The HPB
model is built from the BHZ quantum spin Hall model \cite{BHZ06} by
adding the $\Gamma_B$ term which breaks time-reversal but preserves
inversion: the inversion operation is defined as $\mathcal{P} \equiv
\Gamma_0$ such that $\mathcal{P}H(\vec k) \,\mathcal{P} = H(-\vec k)$.
The authors showed that it has a $\mathbb{Z}_2$ topological index
protected by the $\mathcal{P}$-symmetry.

The HPB model is also $\istar$-symmetric: $\mathcal{I} = \sigma_x
\otimes \tau_z$ such that $\mathcal{I}H(\vec k) \,\mathcal{I} =
H^{*}(\vec k)$. In fact this is the mirror symmetry as noted by the
same authors in Ref.~\onlinecite{hughes-prodan-bernevig11-inversion}.
The $\mathcal{T}$-breaking term $B_x$ plays the role of $\tp$ of the
bilayer model, therefore the model can be cast into the form of
eqn.~\ref{h-blhi} by rotating $(\tau_x, \tau_y)$ to $(\tau_y,
-\tau_x)$. One can then solve it analytically and analyze its Wannier
centers in the same way as the bilayer model. Its $\mathbb{Z}_2$ index is
equivalent to the winding number of $\tilde \gamma$, and is inherited
from the ``monolayer'' (individual spin species) as long as the
central gap does not collapse.

It turns out for this particular model, one can choose to break either
$\mathcal{P}$ or $\istar$ (but not both) and still retain the
nontrivial topology. Here we only consider breaking the $\mathcal{P}$,
for as long as it is preserved, the proof of
Ref.~\onlinecite{hughes-prodan-bernevig11-inversion} applies. To break
$\mathcal{P}$, we add in an $\istar$-symmetric ``spin-flipping''
hopping. The most general form is to replace the $B_x \,\Gamma_B$ term
in eqn.~\ref{hpb-h} by
\begin{gather}
  \begin{pmatrix}
    \mathbf{0} & T \\T\yd & \mathbf{0}
  \end{pmatrix}\quad , \quad T =
  \begin{pmatrix}
    \lambda_1 & \lambda_2 \\
    -\lambda_2 & \lambda_3
  \end{pmatrix}
\end{gather}
where again $\lambda_i$ are complex numbers which could have $\vec k$
dependence. The extra ``$-$'' sign in $T$ as compared with the bilayer
case (eqn.~\ref{t-istar}) will drop after the aformentioned $(\tau_x,
\tau_y) \rightarrow (\tau_y, -\tau_x)$ rotation. One can easily verify
that the condition for inversion symmetry to also hold is $(\lambda_1,
\lambda_2, \lambda_3)_{-\vec k} = (\lambda_1, -\lambda_2,
\lambda_3)_{\vec k}$.

Fig.~\ref{hpbmod} plots the entanglement spectrum and Wilson loop
phases of an $\istar$-symmetric HPB model with broken inversion
symmetry, $m = 1.1$, $(\lambda_1, \lambda_2, \lambda_3) = (0.7, 0.5,
0)$. The topological signatures still persist: the entanglement
quasi-energy spectrum exhibit robust zero modes, and the Wilson loop
phases flow in opposite directions and cross at $\gamma = 0$ and
$\pi$. Since the $\lambda_2$ term explicitly breaks $\mathcal{P}$, the
$k_x$ points where they occur are no longer pinned at
symmetry-invariant points.

\section{Summary}

We have studied the monolayer and bilayer Haldane models and
identified several topological signatures from the real space
perspective. The monolayer zigzag edge modes can be analytically
solved using the {\it Ansatz\/} that the wavefunctions of its $A$ and $B$
sublattices are proportional. This particular form poses restrictions
on the boundary condition which can no longer be prescribed (as
\emph{e.g.} an open boundary) but must be solved self-consistenly, and
the edge state is recognized as an open boundary one when the
``tunnelling strength'' $\rho$ of the two boundaries approaches zero
in the thermodynamic limit. Using the edge solution, the transverse
momentum $k_1 = k_{\rm c}$ at which the two edge modes cross can be
identified as the coplanar point of the $\vec B$ vector that generates
the bulk Hamiltonian $H = \omega + \vec B \cdot \vec \sigma$---that
is, at this point, varying the longitudinal momentum $k_2$ from $0$ to
$2\pi$ will drive $\vec B$ in a closed path on a plane that passes
through the origin. The problem of mapping the $2$-D Brillouin zone
$(k_1, k_2)$ to the sphere $\vec B(k_1, k_2)$ is thus reduced to a
$1$-D mapping from $k_2$ to the ring $\vec B(k_{\rm c}, k_2)$, and the bulk
Chern index reduces to the winding number of the $\vec B$ ring around
the origin.

Interestingly, at $k_{\rm c}$, both the entanglement spectrum and the
Wannier centers exhibit crossings similar to the edge modes. If one
treats the entanglement spectrum as the bipartition coarse graining of
the Wannier centers, then there is a continuous parameterization
$\beta$ such that the eigenvalues $\{r_a\}$ of a band-projected real
space operator $\g R_{\beta} \,\g$ correspond to the entanglement
spectrum at one limit $\beta \rightarrow \infty$, and to the Wannier
centers at the other limit $\beta \rightarrow 0$. The crossing at
$k_{\rm c}$ is universal in that for arbitrary $\beta$, there always exist
$r_a$ levels fixed at $r_a = (N + 1)/2$, where $N$ is the number of
unit cells in the longitudinal direction $\vec a_2$. The special $r_a$
value translates to half occupancy $f = \half$ in the entanglement case,
and to a state localized right in the middle of the $N$-cell chain in
the Wannier case. This universal crossing can be traced back to the
$\vec B$ coplanarity because the two sublattices (after an internal
rotation) are localized in \emph{different} unit cells with the
separation given by the winding number $w$ of the $\vec B$ ring, thus
their average center is a half-odd-integer if $w$ is odd, giving rise
to the special Wannier center, and an entanglement cut placed in
between them will assign the two sublattices to different halves of
the bipartite, yielding an $f = \half$ entanglement occupancy in each
half.

The Haldane bilayer is constructed by Bernal-stacking two
monolayers and allow vertical interlayer hopping $\tp$. Without
exception, the Chern index of bilayer at half filling is the sum of those
of the individual monolayers, which also manifests as the number of
entanglement spectral flows. A special case is when the two monolayers
individually have nonvanishing Chern indices but are opposite to each
other: the nontrivial topology survives in a sense if their parameters
are exactly opposite such that the bilayer bulk Hamiltonian is mapped
to its complex conjugate by inversion, $\mathcal{I}H(\vec
k\mathcal{)}I = H^{*}(\vec k)$ ($\istar$ symmetry): while there is no
gapless edge modes and no entanglement spectral flow, the entanglement
spectrum does exhibit protected $f=\half$ modes. It becomes more
prominent in the Wannier center (Wilson loop phases) spectrum, where
two branches start to \emph{flow} in opposite directions with the
magnitude of their winding number given by the monolayer Chern index.
The flow is robust as long as the central gap of the bilayer does not
collapse when $\tp$ is varied. The role of $\istar$ symmetry is
further confirmed numerically by adding in $\istar$-preserving generic
interlayer hopping. We also found that unlike the monolayer open
boundary edge spectrum whose gapless crossing is a thermodynamic limit
result, the crossing of the opposing Wilson loop phase flows are more
prominent as the number of unit cells $N$ in the $\vec a_2$ direction
is reduced: they are now degenerate for an extended region of $k_1$
instead of crossing at discrete points. This does not happen for the
topologically trivial case where $\istar$ is broken.

The flow-less entanglement spectrum with protected midgap modes is
reminiscent of the $\mathbb{Z}_2$ inversion-symmetric topological
insulators, with the essential difference that the midgap modes there
are pinned at inversion-invariant $k$ points. We looked at one such
model studied in Ref.~\onlinecite{hughes-prodan-bernevig11-inversion}.
This particular model has both inversion ($\mathcal{P}$) and mirror
symmetry. The latter coincides mathematically with the $\istar$
symmetry, rendering the model solvable in the same way as the
$\tp$-only bilayer. The nontrivial topology as captured by the
$\mathbb{Z}_2$ index survives as long as either of $\mathcal{P}$ and
$\istar$ is preserved. The former case falls within the analysis of
Ref.~\onlinecite{hughes-prodan-bernevig11-inversion}. We illustrated
the latter by adding in $\istar$-preserving perturbations that break
$\mathcal{P}$. Such a perturbations shift the $f = \half$ modes and the
$\pi$ Wilson loop phases away from the $\mathcal{P}$-invariant $k$
points but does not generate any avoided crossing.

Both the BHZ quantum spin Hall model and the HPB inversion-symmetric
TI model inherit their $\mathbb{Z}_2$ index from the Chern index of
the underlying single spin species, and from the point of view of the
Wilson loop phases, are topologically equivalent to the situations
where the coupling between the two spin species are turned off,
although for the HPB model this qualitatively changes the edge
spectrum behavior. Mathematically the question becomes how the
interspin coupling (or interlayer coupling in the bilayer case) influences
the flow pattern of the Wilson loop phases, and what kind of coupling
does not degenerate the pattern to the trivial case (\emph{e.g.}, that
of a unity operator). The bilayer Haldane model shows that there are
other types of coupling which break both time-reversal and inversion
and yet still result in a nontrivial topology.

\section{Acknowledgments}
This work was supported by the NSF through grant DMR-1007028.

\appendix
\section{Details of Monolayer Zigzag Edge States}
\label{zzwf}
\subsection{Edge solution}
From \ref{vp-gauge}, we note that as $k_1 \rightarrow -k_1$,
\begin{gather}
  \label{inversion-vp}
  \quad v_1(-k_1) = v_2^{*}(k_1) \quad, \quad p_1(-k_1) = p_1(k_1)
\end{gather}

Eqn.~\ref{ratio} gives two equations (the ratio $r$ itself being yet
undetermined) with two unknowns, $\lambda$ and $\varepsilon$, for each
$k_1$ value. The second equality of eqn.~\ref{ratio} does not involve
$\varepsilon$, and can be used to solve for $\lambda$,
\begin{gather}
  \label{lambda}
  p\nd_2\,v\nd_2\lambda^2 + (v\nd_2 v_1^{*} - v\nd_1 v_2^{*} + p_2^2) \, \lambda +
  p\nd_2\,v_1^{*} = 0,
\end{gather}
This yields eqns.~\ref{lpm} and \ref{rpm}. Note that sending
$\lambda \leftrightarrow \lambda^{-1}$ and $v_1\leftrightarrow
v_2^{*}$ keeps eqn.~\ref{lambda} invariant, thus together with
eqn.~\ref{inversion-vp}, one gets
\begin{gather}
  \label{inversion-lr}
  d(k) = d(-k),\\\lambda_{+}(k) = \lambda_{-}^{-1}(-k), \quad r_{+}(k) =
  r_{-}^{-1}(-k).
\end{gather}

To solve for $\varepsilon$, we use a simple fact about ratios: if $r =
x_1/y_1 = x_2/y_2$, then $r$ is preserved by arbitrary linear
combination of numerators and denominators, $r = (ax_1 + b x_2)/(a y_1
+ b y_2)$, except the unfortunate choice that makes $a x_1 + b
x_2 = 0$. Applying to eqn.~\ref{ratio}, we get
\begin{gather}
  \label{r-e}
  r = {p_2 h_1 - p_2 \varepsilon - 2 p_1 \textsf{Re}\, v_1\over p_2h_2 - p_2
    \varepsilon - 2 p_1 \textsf{Re}\, v_2}
\end{gather}
where $\textsf{Re}$ indicates real part. This gives the two solutions
of $\varepsilon$, eqn.~\ref{e-r}, in terms of $r_{\pm}$ as solved in
eqn.~\ref{rpm}. Similar to eqn.~\ref{inversion-lr},
\begin{gather}
  \varepsilon_{+}(m,k_1) = \varepsilon_{-}(-m, -k_1).
\end{gather}
The bulk spectrum has the same inversion property which is easily
verified from eqn.~\ref{omega-b}.

In the text (eqn.~\ref{discriminant}) we mentioned that $r_{\pm}$ is
real if the discriminant $\Delta$ is non-negative. The reality of
$r_{\pm}$ (eqn.~\ref{discriminant}) in turn has the following
implication: Using $r = r^{*}$ and eqn.~\ref{ratio},
\begin{gather}
  \label{r-lambda}
  r = {v_1^{*} + \lambda p_2\over v_2^{*}} = r^{*} =
 {\lambda^{*}v_1^{*}\over\lambda^{*}v_2^{*} + p_2}
  \implies r = - |\lambda|^2,
\end{gather}
which can also be checked explicitly using eqn.~\ref{rpm}. Notice that
this applies to the singular case, eqn.~\ref{singular}, as well.

That we can get eqn.~\ref{r-lambda} is actually fortunate, for
otherwise there will be no twisted boundary consistent with the
{\it Ansatz\/}. We will come back to this point later when deriving the
eigenstates.

\subsection{Edge crossing point}
To derive the edge crossing point, we note that according to
eqn.~\ref{rpm}, the ratio $r$ depends on the branch: in general $r_{+}
\neq r_{-}$. But eqn.~\ref{r-e} implies that at the edge crossing
point(s), $r_{+} = r_{-}$. It would seem the latter is satisfied only
when the discriminant $\Delta = 0$, but one can easily verify from
Fig.~\ref{ansatz} that these do not correspond to the edge crossing
points. In fact, there is a range of parameters in the topological
phase where $\Delta$ is always positive, \emph{e.g.}, $t_1 = t_2 = t_3
= 0.3$, $m=\phi/\pi = 0.5$. Recall that in deriving eqn.~\ref{r-e}, we
used linear combinations of denominators and numerators of the three
ratios in eqn.~\ref{ratio}, the validity of which requires these
combinations to be non-singular (cf.~discussion leading to
eqn.\ref{r-e}). Thus the only way for eqns.~\ref{rpm} and \ref{r-e} to
be consistent---the former implying $r_{+}\neq r_{-}$, the latter
implying otherwise---is for the linear combinations of both the
denominators and the numerators to be singular,
\emph{viz.}, 
\begin{gather}
  \label{ece}
  p_2 (h_1 - \varepsilon) - 2 p_1 \textsf{Re}\, v_1 = p_2 (h_2 - \varepsilon) - 2
  p_1 \textsf{Re}\,v_2 = 0.
\end{gather}
This yields the edge-crossing condition eqn.~\ref{ec}.

\subsection{Eigenstates}
Eqns.~\ref{edge-eig-eqn-k1} and \ref{edge-eig-eqn-k2} can be cast into
the same Schr\"odinger equation,
\begin{gather}
  \label{sch-eq}
  a \, \psi_{n+1} + b\, \psi_n + c\, \psi_{n-1} = 0\ ,
\end{gather}
where $a$, $b$ and $c$ can either be the set of numerators or
denominators in eqn.~\ref{ratio}, \emph{e.g.}, $a = v_1$, $b = h_1 +
\lambda p_1 - \varepsilon$, $c = v_1^{*} + \lambda\,  p_2$. Note that
these are already known once the edge solutions are obtained.
Eqn.~\ref{sch-eq} is equivalent to
\begin{gather}
  \label{recur}
  \psi_{n+1} - x_1 \psi_n = x_2 \, (\psi_n - x_1 \psi_{n-1})\ ,\\
  x_1 + x_2 = -{b\over a} \quad, \quad x_1\, x_2 = {c\over a}\ ,
\end{gather}
\emph{i.e.}, $x_1$ and $x_2$ are solutions to
\begin{gather}
  a x^2 + b x + c = 0\ .
\end{gather}
Denote $\phi_n = \psi_n - x_1 \psi_{n-1}$, then by eqn \ref{recur},
$\phi_n = x_2^{n-1} \phi_1$, and
\begin{align}
  \notag
  \psi_n &= \phi_n + x_1 \, \psi_{n-1} = \phi_n + x_1 \, \phi_{n-1} + x_1^2
  \, \psi_{n-2}\\
  \notag
  & = \cdots\\
  \notag
  & = \phi_n + x_1 \, \phi_{n-1} + x_1^2 \, \phi_{n-2} + \cdots \\
  & \qquad+ x_1^{n-2} \, \phi_2 + x_1^{n-1}\underbrace{(\phi_1 + x_1
    \psi_0)}_{\psi_1} \ .
\end{align}
\nl (1) If $x_1 \neq x_2$, the geometric series can be summed,
\begin{gather}
  \label{recur-psi}
  \psi_n = {x_1^n - x_2^n\over x_1 - x_2} \> \phi_1 + x_1^n \,\psi_0
  = f_n \,\psi_1 - x_1 \,x_2 \,f_{n-1}\psi_0 \ ,\\
  f_n \equiv {x_1^n - x_2^n\over x_1 - x_2} \ .
\end{gather}
\nl (2) If $x_1 = x_2 = x$, then
\begin{gather}
  \label{recur-x1eqx2}
  \psi_n = n x^{n-1} \psi_1 - (n-1)x^n \psi_0 \ ,
\end{gather}
which is the same as applying L'Hospital's rule on the previous case.
\nl (3) If $ac = 0$, one of the $x$, say $x_2$, is $0$, then
\begin{gather}
  \psi_n = x_1^{n-1} \psi_1 \ .
\end{gather}
\nl Thus in all cases we may proceed with eqn.~\ref{recur-psi}. This
yields
\begin{gather}
  \label{connection}
  \psi\nd_{N+1} = f\nd_{N+1}\psi\nd_1 - x\nd_1 x\nd_2  f\nd_N \psi_0\ ,\\
  \psi\nd_N = f\nd_N \psi\nd_1 - x\nd_1 x\nd_2 f\nd_{N-1} \psi\nd_{0} \ .
\end{gather}

We now come to the issue of boundary conditions. Eqn.~\ref{tbc}
implies
\begin{gather}
  \label{bc-1}
  v\yd
  \begin{pmatrix}
    \psi\nd_0 \\ \lambda\psi\nd_0
  \end{pmatrix}
  = \rho\, v\yd U\yd
  \begin{pmatrix}
    \psi\nd_N \\ \lambda\psi\nd_N
  \end{pmatrix}\ , \\
  \label{bc-2}
  v
  \begin{pmatrix}
    \psi\nd_{N+1} \\ \lambda \psi\nd_{N+1}
  \end{pmatrix}
  = \rho\, U v
  \begin{pmatrix}
    \psi\nd_1 \\ \lambda\psi\nd_1
  \end{pmatrix}\ .
\end{gather}
From eqn.~\ref{vmat},
\begin{gather}
  v^{-1} = {1\over v_1 v_2} 
  \begin{pmatrix*}[r]
    v_2 & 0_{\phantom{1}} \\ -p_2 & v_1
  \end{pmatrix*}
\end{gather}
which exists if $v_1v_2 \neq 0$. The case $v_1 v_2 = 0$ can be
analyzed by Taylor expanding in the vanishing $v_i$ similar to the
discussion leading to eqn.~\ref{singular}. The $v\yd$ can be
dropped from eqn.~\ref{bc-1}. For eqn.~\ref{bc-2}, one gets from
eqn.~\ref{ratio} that
\begin{gather}
  v
  \begin{pmatrix}
    1\\\lambda
  \end{pmatrix} ={v_1\over r}
  \begin{pmatrix}
    r\\\lambda
  \end{pmatrix}\ ,
\end{gather}
thus the two boundary conditions eqns.~\ref{bc-1} and \ref{bc-2} become
\begin{gather}
  \psi_0
  \begin{pmatrix}
    1\\\lambda
  \end{pmatrix}
  = \rho\, \psi_N U\yd
  \begin{pmatrix}
    1\\\lambda
  \end{pmatrix}, \quad 
  \psi_{N+1}
  \begin{pmatrix}
    r\\\lambda
  \end{pmatrix}
  = \rho\, \psi_1 U
  \begin{pmatrix}
    r\\\lambda
  \end{pmatrix}.
\end{gather}
This implies both $\left(
  \begin{smallmatrix}
    1\\\lambda
  \end{smallmatrix}\right)
$ and  $\left(
  \begin{smallmatrix}
    r\\\lambda
  \end{smallmatrix}\right)
$ are eigenstates of $U$,
\begin{gather}
  U\yd
  \begin{pmatrix}
    1\\\lambda
  \end{pmatrix}
  \equiv e^{-i\theta_1}
  \begin{pmatrix}
    1\\\lambda
  \end{pmatrix}, \quad
  U
  \begin{pmatrix}
    r\\\lambda
  \end{pmatrix}
  \equiv e^{i\theta_2}
  \begin{pmatrix}
    r\\\lambda
  \end{pmatrix}.
\end{gather}
Thus, if $U \neq e^{i\theta} \mathbb{I}$, then $\left(
  \begin{smallmatrix}
    1\\\lambda
  \end{smallmatrix}\right)
$ and  $\left(
  \begin{smallmatrix}
    r\\\lambda
  \end{smallmatrix}\right)
$ are either equivalent (if $\theta_1 = \theta_2$), indicating $1 =
r$, or orthogonal (if $\theta_1 \neq \theta_2$), indicating $r =
-|\lambda|^2$. The latter is nothing but eqn.~\ref{r-lambda}. The
boundary conditions then reduce to
\begin{gather}
  \label{su2-bc}
  \psi_0 = \rho\, e^{-i\theta_1}\psi_N\quad , \quad \psi_{N+1} = \rho\,
  e^{i\theta_2}\psi_1\ .
\end{gather}
Substituting these in eqn.~\ref{recur-psi} yields the following
condition,
\begin{multline}
  \label{self-consistent}
  \underbrace{\frac{x_1x_2f_{N-1}}{f_{N+1}}e^{i(\theta_2 - \theta_1)}}_{\equiv A}\rho^2
  \\+ \underbrace{\Bigl[
    \overbrace{\frac{(x_1x_2)^N}{f_{N+1}}}^{\equiv B_1}e^{-i\theta_1}
    + \overbrace{\frac{1}{f_{N+1}}}^{\equiv B_2}e^{i\theta_2}
    \Bigr]}_{\equiv B}\rho - 1 = 0
\end{multline}
where one needs to tune $\theta_1$ and $\theta_2$ such that at least
one solution of $\rho$ is real. A sufficient condition is for both $A$
and $B$ to be real. Here we make $A>0$ so that the solutions of $\rho$
are always real,
\begin{gather}
  \theta_2 - \theta_1 = \arg\left({{f_{N+1}\over x_1x_2f_{N-1}}}\right) \quad, \quad
  A = \left| {x_1x_2 f_{N-1}\over f_{N+1}} \right|,
\end{gather}
then
\begin{gather}
  B = B_1 e^{-i\theta_1} + B_2 e^{i\delta} e^{i\theta_1}
\end{gather}
thus for $B$ to be real,
\begin{gather}
  0 = B - B^{*} = (B_1 - B_2^{*} e^{-i\delta}) e^{-i\theta_1} - (B_1 -
  B_2^{*}e^{-i\delta})^{*} e^{i\theta_1} \\
  \implies \theta_1 = \arg\left(B_1 - B_2^{*}e^{-i\delta}\right) +
  \left\{\begin{matrix}
      0\\\pi
    \end{matrix}\right\}
\end{gather}
where the freedom $\{
\begin{smallmatrix}
  0\\\pi
\end{smallmatrix}\} $ can be used to switch the sign of $B$ (which has
been made real). Then we have
\begin{gather}
  \rho_{\pm} = -{|B|\over 2A} \pm {\sqrt{|B|^2 + 4A}\over 2A}
\end{gather}
Since $|B|$ and $A$ are both $\ge 0$, the $\rho$ with smaller
magnitude (which has a better chance of $\rightarrow 0$ to represent
an open boundary) is always $\rho_{+}$. Notice that each of
$\rho_{\pm}$ still depends on which branch of the edge solution we
have picked in calculating $x_1$ and $x_2$.

In particular, one can show that $\rho = 1$ and $|x_i| = 1$ occur
simultaneously, the former means periodic boundary condition, while
the latter means the {\it Ansatz\/} solution is a bulk solution, so this makes
sense. To see this, assume $\rho = 1$ and $x_2 = e^{i\phi}$, then
eqn.~\ref{self-consistent} becomes
\begin{multline}
  \left[e^{iN\phi}e^{-i\theta_1} - 1 \right] x_1^{N+1} +
  e^{i\phi}\left[ e^{i(\theta_2 - \theta_1)} -
    e^{iN\phi}e^{-i\theta_1} \right] x_1^N \\
  + \left[ e^{i\theta_2} -
    e^{iN\phi}e^{i(\theta_2 - \theta_1)} \right] - \left[
    e^{i\phi}e^{i\theta_2} - e^{i(N+1)\phi}
  \right] = 0\ .
\end{multline}
One then gets $e^{i\theta_1} = e^{i\theta_2} = e^{iN\phi}$ by
requiring all square brackets to be zero, \emph{i.e.},
eqn.~\ref{self-consistent} is indeed consistent.

\subsection{Graphene and boron nitride}
For graphene, $K^{(1)} = K^{(2)} = 0$, so eqns.~\ref{edge-eig-eqn-k1}
and \ref{edge-eig-eqn-k2} become
\begin{gather}
  \lambda R \psi_A = \varepsilon \psi_A\quad , \quad
  R\yd \psi_A = \lambda \varepsilon \psi_A\ ,\\
  R =
  \begin{pmatrix}
    p_1 & & & & zp_2\\
    p_2 & p_1\\
    & p_2 & p_1\\
    & & \ddots & \ddots\\
    & & &p_2 & p_1
  \end{pmatrix}\ ,\\
  p_1 = -2 \cos \frac{k}{2}\quad , \quad p_{2} = -1
\end{gather}
where $z$ controls the boundary condition.

If $\lambda \neq 0$ and $1/\lambda \neq 0$, i.e., both $A$ and $B$
sites have charge density, then $R\,\psi_A = \varepsilon \, \psi_A /
\lambda$ gives
\begin{gather}
  p_2\, \psi_n + p_1\, \psi_{n+1} = {\varepsilon\over\lambda}
  \psi_{n+1}\ , \\
  \psi_0 = z\, \psi_N, \quad n = 0, 1, 2, \ldots, N
\end{gather}
thus
\begin{gather}
  \psi_{n+1} = \left({p_2\over{\varepsilon\over\lambda} -
      p_1}\right)^{\!n+1}\!\! \psi_0 \quad,\quad z =
  \left({p_2\over{\varepsilon\over\lambda} - p_1}\right)^{\!-N}\ ,
\end{gather}
whereas $R\yd \psi_A = \lambda\,\varepsilon \,\psi_A$ will give
\begin{gather}
  p_1 \,\psi_n + p_2\,\psi_{n+1} = \lambda\,\varepsilon\, \psi_n\ ,\\
  \psi_{N+1} = z\, \psi_N, \quad n = 1, 2, \ldots, N, N+1
\end{gather}
yielding
\begin{gather}
  \psi_{n+1} = \left({\lambda \varepsilon -
      p_1\over p_2}\right)^{\!\!n}\psi_1\ ,\ z = \left({\lambda \varepsilon
      - p_1\over p_2}\right)^{\!\!N}\ .
\end{gather}
Equating the expressions for $z$ gives
\begin{gather}
  \lambda = \pm 1 \implies z = \left( {\tilde{\varepsilon}
      - p_1\over p_2} \right)^{\!\!N}
\end{gather}
where $ \tilde \varepsilon \equiv \lambda \varepsilon =
\varepsilon/\lambda$. Equating the two recursions then yields
\begin{gather}
  \tilde\varepsilon = p_1 \pm p_2 = -2 \cos \frac{k}{2} \mp 1, \quad z
  = (\pm 1)^N = \pm 1.
\end{gather}
Since $|z| = 1$, these states correspond to bulk states with periodic
or antiperiodic boundary conditions.

If $\lambda = 0$, then
\begin{gather}
  \begin{pmatrix}
    & R \\ R\yd
  \end{pmatrix}
  \begin{pmatrix}
    \psi_A \\ 0
  \end{pmatrix} = \varepsilon
  \begin{pmatrix}
    \psi_A \\ 0
  \end{pmatrix}
  \\
  \implies \varepsilon = 0\quad , \quad \psi_{n+1} =
  -{p_1\over p_2}\>\psi_n
\end{gather}
and
\begin{gather}
  z = -\left({p_1\over p_2}\right)^{\!\!N} = -(2\cos \frac{k}{2})^N \\
  \implies
  \begin{cases}
    |z| \ll 1 & k \in (\frac{2\pi}{3}, \frac{4\pi}{3})\\
     |z|\gg 1 & k \in (- \frac{2\pi}{3}, \frac{2\pi}{3})\ ,
  \end{cases}
\end{gather}
thus it is an edge solution localized on the $A$ sites at the $y=1$
edge if $|k| > 2\pi/3$ ({\it i.e.\/}, connecting two Dirac points), with zero
energy.

Similarly, if $1/\lambda = 0$, then
\begin{gather}
  \begin{pmatrix}
    & R \\ R\yd
  \end{pmatrix}
  \begin{pmatrix}
    0 \\ \psi_B
  \end{pmatrix} = \varepsilon
  \begin{pmatrix}
    0 \\ \psi_B
  \end{pmatrix} \\
  \implies \varepsilon = 0\quad , \quad \psi_{n+1} = - {p_2\over p_1} \>\psi_n
\end{gather}
and
\begin{gather}
  z = - \left({p_1\over p_2}\right)^{\!\!N} = -(2 \cos \frac{k}{2})^N\\
  \implies 
  \begin{cases}
   |z| \ll 1 & k \in (\frac{2\pi}{3}, \frac{4\pi}{3})\\
   |z| \gg 1 & k \in (- \frac{2\pi}{3}, \frac{2\pi}{3})
  \end{cases}
\end{gather}
thus it is an edge solution localized on the $B$ sites at the $y=N$
edge if $|k| > 2\pi/3$ with zero energy.

For boron nitride (BN), the $A$ sublattice is nitrogen and $B$ is
boron. If $\lambda \ne 0$, $1/\lambda \ne 0$, then $\lambda R \psi_A =
(\varepsilon - m) \psi_A$ yields
\begin{gather}
  \psi_{n+1} = {p_2\over \varepsilon_{-} - p_1} \psi_n\quad , \quad z
  = \left( {\varepsilon_{-} - p_1\over p_2} \right)^{\!\!N}
\end{gather}
where $\varepsilon_{-} \equiv {\varepsilon - m\over \lambda}$.
Similarly $R\yd\psi_A = \lambda (\varepsilon + m) \psi_A$ gives
\begin{gather}
 \psi_{n+1}
  = {\varepsilon_{+} - p_1\over p_2}\quad, \quad z = \left(
   {\varepsilon_{+} - p_1\over p_2} \right)^{\!\!N}
\end{gather}
where $\varepsilon_{+} \equiv \lambda(\varepsilon + m)$. Equating
expressions for $z$ gives
\begin{gather}
  {\varepsilon_{-} - p_1\over p_2} = {\varepsilon_{+} - p_1\over p_2}
\end{gather}
while equating the two $\psi$ recursions gives
\begin{gather}
 {\varepsilon_{-} - p_1\over p_2} ={p_2\over \varepsilon_{+} - p_1}
\end{gather}
thus
\begin{gather}
  {\varepsilon_{-} - p_1\over p_2} ={\varepsilon_{+} -
    p_1\over p_2} = \pm 1 \\
  \implies \varepsilon = \pm \sqrt{m^2 + (p_1 \pm p_2)^2} \quad ,
  \quad |z| = 1\ .
\end{gather}
When $|z| = 1$, these are bulk states.

If $\lambda = 0$, then the edge solution is the same as the
corresponding graphene case with $\varepsilon = m$ and the edge state
is purely on $A$ sites. If $1/\lambda = 0$, then $\varepsilon = -m$
and the edge state is purely on $B$ sites.

\section{Proof of entanglement half occupancy modes for Zigzag edge
  with $t_2 = t_3$}
\label{zzent-zm}
At the edge crossing point, the path of $\vec B$ is coplanar. Rotating
to that plane, the internal-space part of the occupied bulk states are
given by eqn.~\ref{mlh-psi} with $\vartheta = \pi/2$,
\begin{gather}
  |\varphi(k)\rangle = \frac{1}{\sqrt{2}}
  \begin{pmatrix}
    -1\\ e^{i\varphi(k)}
  \end{pmatrix}\ .
\end{gather}
The occupied band projector is thus
\begin{align}
  \g & = \sum_k |k\rangle \langle k | \otimes |\varphi(k)\rangle \langle
  \varphi(k)|\\
  &= \frac{1}{2} \sum_k |k\rangle \langle k| \otimes (\mathbb{I} -
  \sigma_{\varphi(k)})
\end{align}
where $\sigma_{\varphi}$ is the spin operator polarized along the
direction in $xy$ plane with azimuth $\varphi$.
\begin{gather}
  \sigma_{\varphi} \equiv
  \begin{pmatrix}
    0 & e^{-i\varphi}\\ e^{i\varphi} & 0
  \end{pmatrix}\ .
\end{gather}
The restricted correlation matrix is thus
\begin{align}
  G &= \frac{1}{2} \sum_k P_k \otimes (\mathbb{I} -
  \sigma_{\varphi(k)})\\
  &= \frac{\mathbb{I}}{2} - \frac{1}{2}\sum_k P_k \otimes
  \sigma_{\varphi(k)}
\end{align}
where
\begin{gather}
  P_k \equiv R \,|k\rangle \langle k |\, R^{\rm T}
\end{gather}
is the Bloch projector $|k\rangle \langle k|$ restricted to half
space, and $R$ is an oblong matrix to project out the first half,
\begin{gather}
  R =
  \begin{pmatrix}
    1 & & & & & 0\\
    & 1 & & & & & 0\\
    & & \ddots & & & & & \ddots\\
    & & & & 1 & & & & 0
  \end{pmatrix}
\end{gather}
We have used the fact that $\sum_k P_k = \mathbb{I}$ is the
unity in the half system. Instead of the entanglement spectrum, which
is the eigenvalues of $G$, we focus on the spectrum of
\begin{gather}
  \bar G \equiv \mathbb{I} - 2G = \sum_k P_k \otimes
  \sigma_{\varphi(k)}\ .
\end{gather}
An entanglement half occupancy mode $f = \half$ corresponds to a zero
mode of $\bar G$. The order of direct product does not influence the
spectrum, thus $\bar G$ may be written in the following off-diagonal
form,
\begin{gather}
  \bar G =
  \begin{pmatrix}
    & M\\ M\yd
  \end{pmatrix}\quad , \quad M = R \mathcal{M} R^{\rm T}\ ,\\
  \mathcal{M} = \sum_k |k\rangle e^{-i\varphi(k)} \langle k|
\end{gather}

Assume the full system has $2N$ unit cells, then the allowed $k$ are
\begin{gather}
  k_n = {n\pi\over N}\quad , \quad n = 1, 2, \ldots, 2N\ .
\end{gather}
We shall cut the system in equal halves. Let $\bar R$ be the
complement projector to $R$,
\begin{gather}
  \bar R =
  \begin{pmatrix}
    0 & & & & & 1\\
    & 0 & & & & & 1\\
    & & \ddots & & & & & \ddots\\
    & & & & 0 & & & & 1
  \end{pmatrix}\ .
\end{gather}
Then for the real space Bloch waves $|k_n\rangle$,
\begin{gather}
  \bar R \,| k_n\rangle = e^{iN k_n}R \,| k_n\rangle = (-1)^n
  R|k_n\rangle\ .
\end{gather}
Then one may write $\mathcal{M}$ as
\begin{gather}
  \mathcal{M} =
  \begin{pmatrix}
    M & Q\\ Q & M
  \end{pmatrix}
\end{gather}
where $Q = \bar R \mathcal{M} \bar R^{\rm T}$. Then $M$ and $Q$ can be
separated into even and odd parts,
\begin{gather}
  M = M_E + M_O\quad , \quad Q = M_E - M_O
\end{gather}
where
\begin{gather}
  M_E = \sum_{n \text{ even}}R |k_n\rangle e^{-i\varphi(k_n)} \langle
  k_n | R^{\rm T}\\
  M_O = \sum_{n \text{ odd}}R |k_n\rangle e^{-i\varphi(k_n)} \langle
  k_n | R^{\rm T}\ .
\end{gather}
It is also easy to verify that
\begin{gather}
  U\yd \mathcal{M}\, U =
  \begin{pmatrix}
    2 M_E \\ & 2 M_O
  \end{pmatrix}\quad,\quad U = U\yd = \frac{1}{\sqrt{2}}
  \begin{pmatrix}
    \mathbb{I} & \mathbb{I} \\ \mathbb{I} & -\mathbb{I}
  \end{pmatrix}
\end{gather}
thus
\begin{gather}
  \textsf{det}\,(\mathcal{M}) = \textsf{det}\,(4 M_E M_O)
\end{gather}
In fact, the two sets $\{R|k_n\rangle\}$ with even and odd $n$ are
both complete sets in the half space, but normalized to $1/\sqrt{2}$
due to the projection, thus $2M_E$ and $2M_O$ are both unitary,
\begin{gather}
  \label{mmd}
  M_E M_E\yd = M_O M_O\yd = \frac{\mathbb{I}}{4}
\end{gather}
and
\begin{gather}
  \label{detm}
  \textsf{det}\,(2 M_E) = e^{-i \Phi_E}\quad , \quad \textsf{det}\,(2 M_O) = e^{-i\Phi_O}
\end{gather}
where
\begin{gather}
  \Phi_E = \sum_{n \text{ even}} \varphi(k_n)\quad , \quad \Phi_O =
  \sum_{n \text{ odd}} \varphi(k_n)
\end{gather}

Now, when $t_2 = t_3$, both $v_1$ and $v_2$ are real
(eqn.~\ref{vp-gauge}), so at $k_1 = k_{\rm c}$, changing $k_2 \rightarrow -
k_2$ only changes the sign of $B_y$ (eqn.~\ref{bkc}). Thus in the
rotated $\vec B$ plane, $\varphi(-k) = - \varphi(k)$. In particular,
\begin{gather}
  \varphi(\pi) = 0 \text{ or }\pi
\end{gather}
and $\varphi(0)$ is defined as $0$. $M_E$ and $M_O$ are both real, and
\begin{gather}
  \textsf{det}\,(4 M_E M_O) = e^{-i(\Phi_E + \Phi_O)} = e^{-i\varphi(\pi)} = \pm 1
\end{gather}
because all other $\varphi(k)$ are cancelled by $\varphi(-k)$. Then
\begin{align}
  \textsf{det}\, M &= \textsf{det}\,(M_E + M_O) = {\textsf{det}\, (M_E M_E^{\rm T} + M_O M_E^{\rm T})\over \textsf{det}\,
    M_E}\\
  & = {\textsf{det}\, (M_O M_O^{\rm T} + M_O M_E^{\rm T})\over \textsf{det}\, M_E} \\
  &= {\textsf{det}\, M_O\over \textsf{det}\, M_E} \textsf{det}\, M^{\rm T} =
  \begin{cases}
    + \textsf{det}\, M^{\rm T} & \text{if } \varphi(\pi) = 0\\
    - \textsf{det}\, M^{\rm T} & \text{if } \varphi(\pi) = \pi
  \end{cases}
\end{align}
where in the first line, we multiplied and divided by $\textsf{det}\, M_E^{\rm T}$,
and in obtaining the second line, we replaced $M_EM_E^{\rm T}$ with
$M_OM_O^{\rm T}$ which follows from eqn.~\ref{mmd}. Thus
\begin{gather}
  \textsf{det}\, \bar{G} = (\textsf{det}\, M)^2 = 0 \text{ if } \varphi(\pi) = \pi\ ,
\end{gather}
\emph{i.e.}, there exist entanglement half occupancy modes.

\section{Non-Abelian Wannier centers and Wilson loops}
\label{wcwl}
\newcommand{\hk}{H}\newcommand{\hfull}{\mathcal{H}}%

Consider, for simplicity, a periodic $1$-D system. Higher-dimensional
systems can be analyzed by parameterizing the system with momenta
$(\kpara, \kperp)$ and considering the effectively $1$-D system at
fixed $\kperp$. Assume there are $q$ bands and $N$ unit cells. The
full Hamiltonian is a $qN\times qN$ matrix $\hfull$ and its Fourier
transform is a $q\times q$ matrix $\hk(k)$. The eigenstates of
$\hfull$ are $qN$-component Bloch states $|\Psi_{k_n}^a\rangle$,
\begin{gather}
  |\Psi_{k_n}^a\rangle = |k_n\rangle \otimes |\psi_{k_n}^a\rangle\quad,\quad a = 1, 2, \ldots, q\\
  k_n = {2\pi n\over N}\quad,\quad n = 1, 2, \ldots, N \ .
\end{gather}
The $q$-component $|\psi_{k_n}^a\rangle$ is the $a^{\rm th}$ band
eigenstate of $\hk(k)$, and the $N$-component $|k_n\rangle$ is the
Bloch phase, $\langle x | k_n\rangle = \exp(i k_n x)/\sqrt{N}$. The
Wannier states can be defined as eigenstates of $\g
R\,\g$\cite{Yu11-z2-wannier}, where
\begin{gather}
  \g = \sum_{a = 1}^{\nu} \sum_{n = 1}^N |\Psi_{k_n}^a\rangle\langle \Psi_{k_n}^a|
\end{gather}
is the projector onto the $\nu$ occupied bands, and
\newcommand{\dk}{\delta k}%
\begin{gather}
  R = \exp(i \hat X \dk) \otimes \mathbb{I}\ .
\end{gather}
$\hat X$ is the position operator in real-space (\emph{i.e.},
measuring unit cell coordinates), and $\dk = 2\pi/N$ is the step in
the discrete $\{k_n\}$. Note that $\exp(i\hat X \dk)$ is the momentum
space translation operator,
\begin{gather}
  \exp(i\hat X \dk) | k_n\rangle = |k_n + \dk\rangle =
  |k_{n+1}\rangle\ ,
\end{gather}
this ensures the resulting Wannier states to have proper translational
symmetry in real space. The eigenstates of $\g R \g$ belong to the
occupied bands and thus have the decomposition,
\begin{gather}
  |\Phi_{\lambda}\rangle = \sum_{a = 1}^{\nu} \sum_{n=1}^N f_{k_n}^a
  |\Psi_{k_n}^a\rangle\quad , \quad \g R \,\g \, |\Phi_{\lambda}\rangle =
  \lambda |\Phi_{\lambda}\rangle\ .
\end{gather}

To solve for $|\Phi_{\lambda}\rangle$, it is useful to introduce the
$\nu \times \nu$ overlap matrix $\hat U_{mn}$ between the
internal-space bases at different $k_m$ and $k_n$ points,
\begin{gather}
  \bigl[\hat U_{mn}\bigr]_{ab} = \langle \psi_{k_m}^a |
  \psi_{k_n}^b\rangle\,,\, a,b = 1, 2, \ldots, \nu
\end{gather}
from which one can define
\begin{gather}
  \label{u-cumu}
  \uu(m \leftarrow n) = \hat U_{m\,m-1} \hat U_{{m-1\,m-2}} \cdots \hat U_{n+1\,n}\quad ,\\
  \ww = \uu(N \leftarrow 0)\,,
\end{gather}
which are also $\nu \times \nu$ matrices. Furthermore, denote
$\{f_{k_n}^a\}$ collectively as a $\nu$-component column vector
\begin{gather}
  \fvec_n = (f_{k_n}^1, f_{k_n}^2, \ldots, f_{k_n}^{\nu})^{\rm T} \ .
\end{gather}
Periodicity in the $n$ index is understood, \emph{viz.}, $\fvec_0 =
\fvec_N$, $\hat U_{10} = \hat U_{1N}$. The eigenvalue problem can now
be cast into
\begin{gather}
  \label{u-recursion}
  \hat U_{n\,n-1} \fvec_{n-1} = \lambda \, \fvec_n \implies
  \ww\, \fvec_N = \lambda^N \fvec_N\,.
\end{gather}
One then diagonalizes $\ww$ with eigenvectors $\{\mathbf{g}_w\}$,
\begin{gather}
  \ww \,\mathbf{g}_w = \rho_w \, e^{i \gamma_w}\ \mathbf{g}_w\ ,\ w = 1,
  2, \ldots, \nu
\end{gather}
and obtain the solutions to eqn.~\ref{u-recursion},
\begin{gather}
  \label{lwi}
  \lambda_{w,I} = (\rho_w)^{\dk/2\pi}
  \exp\Big\{i\bigl(\frac{\gamma_w}{2\pi} + I\bigr)\dk\Big\}\,,\\
  \label{fwi}
  \fvec_n(w,I) = {\uu(n\leftarrow 0)\over (\lambda_{w,I})^n}\>
  \mathbf{g}_w
\end{gather}
where the integer $I = 0, 1, \ldots N-1$ labels the unit cell around
which the Wannier states $|\Phi_{\lambda_{w,I}}\rangle$ are localized.

In the continuum limit $N \rightarrow \infty$, the basis-overlap
matrix $\hat U_{n\,n-1}$ is related to the non-Abelian Berry
connection matrix $\hat A$ via
\begin{align}
  \bigl[\hat U_{n\,n-1}\bigr]_{ab} &= \delta_{ab} + i \,\dk \,
  \langle\psi_{k_n}^a |\, i\, \partial_k \,| \psi_{k_n}^b\rangle \notag \\
  &=\bigl[e^{i \dk \hat A(k_n)}\bigr]_{ab}
\end{align}
where $\hat A_{ab} = \langle \psi_k^a |\, i\,\partial_k \,| \psi_k^b\rangle$
, hence
\begin{gather}
  \uu(m \leftarrow n) \rightarrow \wwp \exp \left\{i
    \int\limits_{k_n}^{k_m} dk\, \hat A(k)\right\}\,,\\
  \ww \rightarrow \wwp \exp\left\{i\int \limits_0^{2\pi} dk\, \hat A(k)
  \right\}
\end{gather}
where $\wwp$ denotes path ordering. $\ww$ is nothing but the Wilson
loop operator, and is now unitary, \emph{i.e.}, $\rho_w \rightarrow
1$. $\g \bigl[\hat X\otimes \mathbb{I}\bigr] \g$ shares the same
eigenstates with $\g R \g$, with its eigenvalues given by the phases
in eqn.~\ref{lwi}
\begin{gather}
  x_{w,I} = {\gamma_w\over 2\pi} + I
\end{gather}
This agrees with the continuum limit result\cite{Qi11-wannier}. We
shall take this as the Wannier centers even when $N$ is finite.

Eqn.~\ref{u-cumu} would seem to suggest that the Wilson loop phases
depend on the phase convention of $\{|\psi_{k_n}^a\rangle\}$ at all
$k_n$, but this is not true. Notice that
\begin{gather}
  \bigl[\hat U_{mn} \hat U_{np}\bigr]_{ab} = \langle \psi_{k_m}^a |
  P\nd_{k_n} | \psi_{k_p}^b\rangle
\end{gather}
where $ P_{k_n} = \sum_{c = 1}^{\nu} |\psi_{k_n}^c\rangle \langle
\psi_{k_n}^c|$ is the internal space projector onto occupied bands at
$k_n$, and does not depend on the phase convention. Thus 
\begin{gather}
  \bigl[\uu(m\leftarrow n)\bigr]_{ab} = \langle \psi_{k_m}^a | \hat{\mathbf{U}}(m\leftarrow n) | \psi_{k_n}^b\rangle \,,\\
  \hat{\mathbf{U}}(m \leftarrow n) \equiv P\nd_{k_m} P\nd_{k_{m-1}}
  P\nd_{k_{m-2}} \cdots P\nd_{k_{n+1}} P\nd_{k_n}
\end{gather}
$\hat{\mathbf{U}}$ can be understood as a $q \times q$ matrix in the
sublattice basis (recall that $q$ is the total number of bands).
Correspondingly, 
\begin{gather}
  \label{monodromy-app}
  \wwb = P\nd_{k_N} P\nd_{k_{N-1}} P\nd_{k_{N-2}} \cdots
  P\nd_{k_2} P\nd_{k_1} P\nd_{k_N}\,.
\end{gather}
and $\ww$ is a $\nu \times \nu$ submatrix of
$V\yd\wwb V$ in the occupied-band block at $k_N$,
where $V$ diagonalizes $H(k_1, k_2 = 2\pi)$. In fact
\begin{gather}
  \label{wsub}
  V\yd\,\wwb \,V =
  \begin{pmatrix}
    \ww & \mathbf{0}\\
    \mathbf{0} & \mathbf{0}
  \end{pmatrix}
  \begin{matrix}
    \leftarrow & \nu \text{ occupied bands}\\
    \leftarrow & q-\nu \text{ empty bands}
  \end{matrix}
\end{gather}
since any matrix element involving the empty bands are projected out
by the head and tail $P_{k_N}$. Thus $\wwb$ and $\ww$ have
the same non-zero eigenvalues.

\section{Invariance of eigenvalues of product of projectors under cyclic permutation}
\label{pprod}
Assume
\begin{gather}
  \label{p1-n}
  P_1P_2 \cdots P_N \,| \psi\rangle = \lambda\, |\psi\rangle\quad,\quad\lambda \neq 0
\end{gather}
where $\{P_n\}$ is a set of arbitrary projectors (square matrices
zero-padded to a common dimension if necessary). Now apply the tail
projector to both sides, one gets that
\begin{gather}
  P_N P_1 P_2 \cdots P_{N-1} (P_N | \psi\rangle) = \lambda (P_N |
  \psi\rangle)\,,
\end{gather}
\emph{i.e.}, $P_N \,| \psi\rangle$ is an eigenvector of the permuted
product $P_N P_1 P_2\cdots P_{N-1}$ with the same eigenvalue. Carrying
out the procedure recursively, we have that
\begin{gather}
  P_n P_{n+1}\cdots P_N P_1 P_2 \cdots P_n \, | \psi_n\rangle = \lambda | \psi_n\rangle\,,\\
  |\psi_n\rangle \equiv P_n P_{n+1} \cdots P_N \, | \psi\rangle\,,
\end{gather}
\emph{i.e.}, $|\psi_n\rangle$ is an eigenstate with the same
eigenvalue $\lambda$ after cyclicly permuting the product for $n$
times. Since all non-zero eigenvalues are preserved and cyclic
permutation does not change dimension, we have that all eigenvalues
are preserved.

A corollary is that for any two projectors $P$ and $R$, $PRP$ and
$RPR$ have the same spectrum, because $\textsf{spec}(PRP) =
\textsf{spec}(RPP) = \textsf{spec}(RRP) = \textsf{spec}(RPR)$.

\bibliographystyle{apsrev-no-url}
\bibliography{mblh}
\end{document}